\documentclass[aip,jcp,11pt,reprint]{revtex4-1}
\usepackage[utf8]{inputenc}
\usepackage{amsmath}
\usepackage{color}
\usepackage{graphicx}

\newcommand{\ket}[1]{\left| #1 \right>} 
\newcommand{\bra}[1]{\left< #1 \right|} 
\newcommand{\bv}[1]{\ensuremath{\mathbf{#1}}} 
\newcommand{\avg}[1]{\left< #1 \right>} 

\begin{document}
\title{Time-dependent coupled cluster theory on the Keldysh contour for non-equilibrium systems}
\date{July 2019}
\author{Alec F. White}
\email{whiteaf@berkeley.edu}
\author{Garnet Kin-Lic Chan}
\email{gkc1000@gmail.com}
\affiliation{Division of Chemistry and Chemical Engineering, California Institute of Technology, Pasadena,
California 91125, USA}

\begin{abstract}
We leverage the Keldysh formalism to extend our implementation of finite temperature coupled cluster theory [\textit{J. Chem. Theory Comput.} 2018, \textit{14}, 5690-5700] to thermal systems that have been driven out of equilibrium. The resulting Keldysh coupled cluster theory is discussed in detail. We describe the implementation of the equations necessary to perform Keldysh coupled cluster singles and doubles calculations of finite temperature dynamics, and we apply the method to some simple systems including a Hubbard model with a Peierls phase and an {\it ab initio} model of warm-dense silicon subject to an ultrafact XUV pulse.
\end{abstract}
\maketitle

\section{Introduction}\label{sec:intro}
Understanding the behavior of condensed phase systems driven out of equilibrium is key to the development of technologies
that can make use of exotic non-equilibrium electronic properties  in real devices. Ultrafast spectroscopy provides the experimental means to probe non-equilibrium dynamics, and  has been used to study transient electronic phases of materials.\cite{Kubler2007,Lu2008,Fausti2011,Kim2012,Morrison2014}
However, the direct simulation of the dynamics is complicated by several factors including:
\begin{enumerate}
    \item The quantum many-body problem for a realistic material
    \item The statistical mixture of many-body states implied by a finite temperature
    \item The generically non-thermal, time-dependent distribution of states generated by the perturbation
    \item The coupling between electronic and nuclear degrees of freedom.
\end{enumerate}

For simplicity in this study we will neglect the description of electron-nuclear coupling.
However, points 1.-3. already pose major challenges for an electronic treatment.
In {\it ab initio} real time dynamics, 
there have been applications of real-time density functional theory (DFT) to both
materials as well as molecules\cite{Gross2004,Botti2007,Otobe2009,Burgdorfer2014,Sato2015,Ullrich1998,Tong2001,Baer2003,Marques2003,Cheng2006},
while time-dependent {\it ab initio} wavefunction methods such as time-dependent configuration interaction~\cite{Krause2005,Schlegel2007,Greenman2010} and coupled cluster methods have been introduced in molecules\cite{Schonhammer1978,Hoodbhoy1978,Hoodbhoy1979,Huber2011,Kvaal2012,Nascimento2017}.
These works however, have typically neglected temperature either due to the complications introduced, or because it is not of interest
to the phenomenon studied. Finite-temperature electron dynamics has been studied using model Hamiltonians with a variety of methods,
such as time-dependent exact diagonalization\cite{Chen2009}, time-dependent density matrix renormalization group methods\cite{Cazalilla2002,Verstraete2004,Kokalj2009,Wolf2014,Ren2018}, diagrammatic Monte Carlo\cite{Werner2009,Schiro2009,Segal2010,Werner2010,Antipov2017}, and hierarchical
equations-of-motion\cite{Jin2008,Huo2011,Schinabeck2018}. However, the {\it ab initio} extensions of such
studies remain to be developed.

Here, we present a method to model finite-temperature electronic dynamics that
is suited to {\it ab initio} simulations of condensed phase systems as well as molecules.
We work within the coupled cluster framework, which has enjoyed great popularity for solving the zero-temperature electronic structure problem,\cite{Cizek1980,Bartlett1981,Crawford2000,Shavitt2009} and where extensions to periodic solids\cite{Hirata2001,Hirata2004,Booth2013,McClain2017}, finite temperatures\cite{Sanyal1992,Mandal2003,White2018,Hummel2018}, and open systems~\cite{Dzhioev2015} have become areas of recent research.
From a theoretical perspective, one can either start from the equation-of-motion for the density matrix or from a thermofield formalism  which propagates a pure state in an extended Hilbert space\cite{Seminoff1983,Matsumoto1983,Matsumoto1985}.
We work within the density matrix language, and use the Keldysh formalism\cite{Schwinger1961,Kadanoff1962,Keldysh1965}
to provide a straightforward generalization of the imaginary-time finite-temperature CC equations in Ref.~\onlinecite{White2018}
to describe the dynamics of finite temperature systems.
We show some initial applications to model systems and a simple {\it ab initio} representation of high-temperature silicon under the influence of an ultrafast extreme ultraviolet (XUV) pulse.



\section{Theory}
We will describe the theory in 4 steps:
\begin{enumerate}
    \item In Section~\ref{sec:FTCC} we describe the working equations of the FT-CC method presented in Ref.~\onlinecite{White2018}
    \item In Section~\ref{sec:Keldysh} we describe some important aspects of the Keldysh formalism
    \item In Section~\ref{sec:Keldysh_CC} we generalize the imaginary time FT-CC to out of equilibrium systems using the Keldysh formalism
    \item In the final Sections (\ref{sec:KeldyshCC_cor}-\ref{sec:KelConserv}) we discuss several important properties and extensions of the theory.
\end{enumerate}

\subsection{Finite temperature coupled cluster}\label{sec:FTCC}
In Ref.~\onlinecite{White2018} we presented a finite temperature coupled cluster theory which provides an ansatz for the correlation contribution to the grand potential $\Omega$ such that
\begin{equation}
    \Omega = \Omega^{(0)} + \Omega^{(1)} + \Omega_{CC}
\end{equation}
where $\Omega^{(0)} + \Omega^{(1)}$ is the grand potential of the thermal Hartree-Fock theory and $\Omega_{CC}$ is the approximation
computed by coupled cluster.
Our strategy was to generalize the diagrammatic rules of finite temperature perturbation theory to the diagrammatic representation of the coupled cluster iteration. Interestingly, this statregy ultimately led to equations identical to those of Mukherjee's thermal cluster cumulant theory (TCC)\cite{Sanyal1992,Sanyal1993,Mandal2001,Mandal2003} (the relationship between the thermal Wick expansion used in TCC and thermofield dynamics is also
discussed in Mukherjee's work~\cite{Sanyal1993}).
The coupled cluster correction to the grand potential is computed from time-dependent amplitudes, $s_{\mu}(\tau)$, evaluated in imaginary time:
\begin{eqnarray}\label{eqn:FT_CC_Omega}
	\Omega_{CC} &=& \frac{1}{4\beta}\sum_{ijab}\langle ij||ab\rangle
	\int_0^{\beta}d\tau
	[s_{ij}^{ab}(\tau) + 2s_i^a(\tau)s_j^b(\tau)] \nonumber \\
	&+& \frac{1}{\beta} \sum_{ia}f_{ia}\int_0^{\beta}d\tau s_i^a(\tau).
\end{eqnarray}
Here, $\beta$ is the inverse temperature,  we use $f$ and $\langle pq || rs\rangle$ as the 1-particle and antisymmetrized 2-particle integrals of the interaction, and the sums run over all orbitals. The set of generalized excitation operators enumerated by $\mu$ defines the truncation of the theory, and the amplitudes are determined from a set of non-linear Volterra integral equations of the form
\begin{equation}\label{eqn:ftcc_ampl}
    s_{\mu}(\tau) = - \int_0^{\tau}d\tau' e^{\Delta_{\mu}(\tau' - \tau)}\text{S}_{\mu}[\bv{s}(\tau')].
\end{equation}
$\Delta_{\mu}$ denotes a difference of orbital energies, and the kernel, S$_{\mu}$, is a nonlinear function of the amplitudes that is local in imaginary time. It contains contractions that are identical in graphical representation to those of ground state coupled cluster theory, the only difference algebraically being the inclusion of Fermi-Dirac occupation numbers and sums that run over all orbitals. Explicit expressions for the FT-CCSD S$_1$ and S$_2$ kernels are given in the appendix of Ref.~\onlinecite{White2018}.

Properties of the thermal system can then be computed by differentiation of $\Omega$. In practice, this is accomplished by defining a Lagrangian, 
\begin{widetext}
\begin{equation}\label{eqn:Lagrangian}
	\mathcal{L} \equiv \frac{1}{\beta}\int_0^{\beta}\mathrm{E}(\tau) -
	\frac{1}{\beta}\sum_{\mu}\int_0^{\beta} d\tau \lambda_{\mu}(\tau)\left[
	s_{\mu}(\tau) + \int_0^{\tau}d\tau' e^{\Delta_{\mu}(\tau' - \tau)} \mathrm{S}_{\mu}(\tau')\right].
\end{equation}
\end{widetext}
(Equation 37 of Ref.~\onlinecite{White2018}) with Lagrange multipliers $\lambda_{\mu}(\tau)$ such that $\mathcal{L}$ coincides with the free energy at its stationary point but is a variational function of $\lambda_{\mu}(\tau)$ and $s_{\mu}(\tau)$. Variational optimization of $\mathcal{L}$ with respect to $\lambda_{\mu}$ yields the amplitude equations (Equation~\ref{eqn:ftcc_ampl}) and variational optimization with respect to $s_{\mu}$ yields linear equations for $\lambda_{\mu}$ of the form 
\begin{equation}
    \lambda_{\mu}(\tau) = - \text{L}_{\mu}[\bv{s}(\tau),\tilde{\bv{\lambda}}(\tau)].
\end{equation}
The L$_{\mu}$ kernel is also local in imaginary time and contains the same contractions as the ground state $\lambda$ equations. We use the notation given by
\begin{equation}
    \tilde{\lambda}_{\mu}(\tau) = \int_\tau^{\beta}d\tau'e^{\Delta_{\mu}(\tau - \tau')}\lambda_{\mu}(\tau').
\end{equation}
The discretized versions of these equations are given in Appendices A and C of Ref.~\onlinecite{White2018}.

\subsection{The Keldysh formalism for non-equilibrium systems}\label{sec:Keldysh}
Our approach to the non-equilibrium problem is based on the Keldysh formalism.\cite{Schwinger1961,Kadanoff1962,Keldysh1965,VanLeeuwen2006} This approach is conceptually appealing because the 3 generic difficulties mentioned in Section~\ref{sec:intro} are treated on the same footing. Many-body correlation, initial finite-temperature, and the dynamics of a driven system are all treated naturally in a single time-dependent framework. We introduce the basic ideas and notations in Appendix~\ref{sec:AKeldysh}. 

The key to the Keldysh formalism is the compact representation of the out-of-equilibrium average of some property as a trace of a contour-ordered exponential (see Equation~\ref{eqn:KelyshAvg}). In recent years, many-body methods using this formalism have gone by the name ``Kadanoff-Baym dynamics"\cite{Kwong2000,VonFriesen2009,PuigVonFriesen2010}. Such methods commonly compute the Green's functions along the Keldysh contour,\cite{Kwong2000,K.2013,Aoki2014,Spicka2014,Do2014} but we will instead use the non-equilibrium analog of the free energy to compute properties. We will refer to this quantity as the Keldysh cumulant generating function.

\begin{figure}
    \centering
    \includegraphics[scale=1.0]{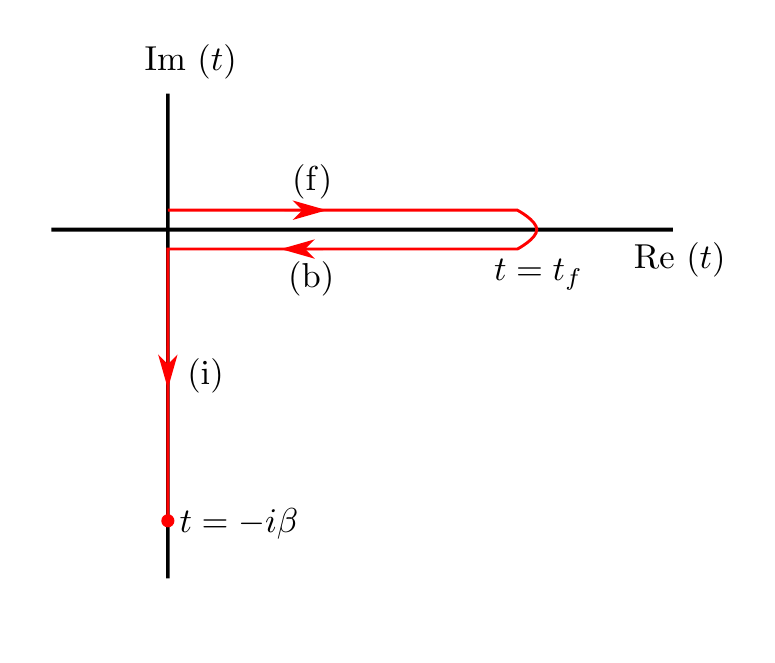}
    \caption{A cartoon depiction of the Keldysh contour in the complex time plane with forward (f), backward (b), and imaginary (i) branches labelled. Note that we draw the forward (backward) branches some small amount above (below) the real axis for clarity. In actual calculations, the forward and backward branches lie on the real axis.}
    \label{fig:contour}
\end{figure}

Consider the grand potential represented as the logarithm of the trace of a time-ordered exponential:
\begin{equation}
	\Omega = -\frac{1}{\beta}\ln\text{Tr}\left[e^{-\beta K_0}\mathcal{T}
	\exp\left(-i\int_{0}^{-i\beta} dt V_I(t)\right) \right].
\end{equation}
$K_0$ is the zeroth order Hamiltonian including the chemical potential (Equation~\ref{eqn:K0}), and $V_I(t)$ is the interaction in the interaction picture.
We can rewrite this expression as an integral along the Keldysh contour $C$ because the contribution from the forward ($0\rightarrow t_f$) and backward ($t_f \rightarrow 0$) branches will exactly cancel (see Fig.~\ref{fig:contour}):
\begin{equation}
	\Omega = -\frac{1}{\beta}\ln\text{Tr}\left[e^{-\beta K_0}\mathcal{T}_C
	\exp\left(-i\int_C dt V_I(t)\right) \right].
\end{equation}
We may now introduce a functional dependence on a parameter $\alpha$ such that
\begin{equation}
	\Omega[\alpha] = -\frac{1}{\beta}\ln\text{Tr}\left[e^{-\beta K_0}\mathcal{T}_C
	\exp\left(\alpha O(t_f)-i\int_C dt V_I(t)\right) \right].
\end{equation}
Notice that for $\alpha = 0$ this is just the equilibrium grand potential,
\begin{equation}
	\Omega[0] = \Omega,
\end{equation}
but the derivative of $\Omega[\alpha]$ evaluated $\alpha = 0$ will reproduce Equation~\ref{eqn:KelyshAvg} with an extra factor of $\beta$:
\begin{align}
	\left.\frac{\partial\Omega}{\partial\alpha}\right|_{\alpha = 0} &= -\frac{1}{\beta Z}
	\text{Tr}\bigg\{e^{-\beta K_0} \nonumber \\
	&\times
	\mathcal{T}_C\left[\exp\left(-i\int_C dt V_I(t)\right)O(t_f)\right]
	\bigg\}.
\end{align}
This means that the derivatives of $\Omega[\alpha]$, the Keldysh cumulant generating function, are the observables of the non-equilibrium system
\begin{equation}
	\avg{O(t_f)} = -\beta\left. \frac{\partial\Omega[\alpha]}{\partial\alpha}\right|_{\alpha = 0}
\end{equation}

\subsection{Non-equilibrium properties from Keldysh coupled cluster}\label{sec:Keldysh_CC}
In order to compute this quantity as a straightforward extension of finite temperature coupled cluster, we slightly rewrite $\Omega[\alpha]$ as
\begin{widetext}
\begin{equation}
	\Omega[\alpha] = -\frac{1}{\beta}\ln\text{Tr}\left[e^{-\beta K_0}\mathcal{T}_C
	\exp\left(-i\int_C dt [V_I(t) + i\alpha O(t)\delta_C(t - t_f)]\right) \right].
\end{equation}
\end{widetext}
The expression is now analogous to what we compute in the equilibrium version of finite temperature coupled cluster. The only difference is that we have a modified potential,
\begin{equation}
	V_I(t) + i\alpha O(t)\delta_C(t - t_f),
\end{equation}
where $\delta_C$ indicates that the delta function acts only once along the contour. In practice there is great flexibility in how the contour is chosen as long as it passes through $t_f$ at least once.

FT-CC provides an approximation to $\Omega$ by imaginary time-integration of non-linear amplitude equations. We now extend this same approximation to systems out of equilibrium by solving the same non-linear amplitude equations along the Keldysh contour. This leads to complex amplitudes $s_{\mu}(t)$ where the time argument, $t$, runs along the Keldysh contour. The coupled cluster contribution to the grand potential is given by
\begin{eqnarray}\label{eqn:K_CC_Omega}
	\Omega_{CC} &=& \frac{i}{4\beta}\sum_{ijab}\langle ij||ab\rangle
	\int_Cdt
	[s_{ij}^{ab}(t) + 2s_i^a(t)s_j^b(t)] \nonumber \\
	&+& \frac{i}{\beta} \sum_{ia}f_{ia}\int_C dt s_i^a(t).
\end{eqnarray}
For computational reasons, the integrals in the amplitude equations are broken up so that
\begin{widetext}
\begin{align}
    s_{\mu}^f(t) &= -i\int_0^{t}dt' e^{i\Delta_{\mu}(t' - t)}\text{S}_{\mu}[\bv{s}^f(t')] \\
    s_{\mu}^b(t) &= e^{i\Delta_{\mu}(t_f - t)}s_{\mu}^f(t_f)-i\int_{t_f}^t dt' e^{i\Delta_{\mu}(t' - t)}\text{S}_{\mu}[\bv{s}^b(t')] \\
    s_{\mu}^i(\tau) &= s_{\mu}^b(0)e^{-\Delta_{\mu}\tau} -
    \int_0^{\tau}d\tau'e^{\Delta_{\mu}(\tau' - \tau)}
    \text{S}_{\mu}[\bv{s}^i(\tau')]
    \label{eqn:amplitude_eqns}
\end{align}
\end{widetext}
where we have used the superscripts $f,b,i$ to refer to the forward, backward, and imaginary parts of the Keldysh contour (see Figure~\ref{fig:contour}). Explicit expressions for the kernel, $\mathrm{S}_{\mu}$, of Keldysh-CCSD are given in Appendix~\ref{sec:aampl}.

We now define a Lagrangian in analogy with Equation~\ref{eqn:Lagrangian}:
\begin{widetext}
\begin{equation}\label{eqn:KLagrangian}
	\mathcal{L} \equiv \frac{i}{\beta}\int_Cdt\mathrm{E}(t) -
	\frac{i}{\beta}\int_C dt \lambda^{\mu}(t)\left[
	s_{\mu}(t) + \int_{C(0)}^{C(t)}dt' e^{i\Delta_{\mu}(t' - t)} \mathrm{S}_{\mu}(t')\right].
\end{equation}
\end{widetext}
We have used $C(t)$ to indicate the limits of the integral along the Keldysh contour. In Appendix~\ref{sec:Alambda}, we discuss the equations determining the multipliers, $\lambda^{\mu}(t)$, along the Keldysh contour. From the amplitudes and Lagrange multipliers, we can construct (unrelaxed) coupled cluster density matrices at each time along the contour as described in Appendix~\ref{sec:A1rdm}. This means that once we solve the CC equations on the Keldysh contour, we can compute any 1-particle property for any real time $t\leq t_f$. Two particle properties can also be computed, though the additional computational cost is significant. Some details of the implementation are discussed in Section~\ref{sec:Implementation}.

\subsection{Correlation functions from higher-order response}\label{sec:KeldyshCC_cor}
The theory can also be extended to the computation of correlation functions,
\begin{equation}
    \avg{O_1(t_1)O_2(t_2),\ldots}
\end{equation}
in or out of time order. We must consider a more general Keldysh cumulant generating function
\begin{widetext}
\begin{equation}
    \Omega[\alpha_1,\alpha_2,\ldots] = -\frac{1}{\beta}\ln\text{Tr}\left[e^{-\beta K_0}\mathcal{T}_C
	\exp\left(-i\int_C dt [V_I(t) + i\alpha_1 O_1(t)\delta_C(t - t_1) + i\alpha_2 O_2(t)\delta_C(t - t_2) + \ldots]\right) \right]
\end{equation}
\end{widetext} 
and use the flexibility of the Keldysh formalism to choose the contour and the locations of the contour delta functions ($\delta_C$) such that contour ordering reproduces the ordering of the operators in the desired correlation function. One can then compute the desired $n$-point correlation function by the $n$th order derivative:
\begin{equation}
    \avg{O_1(t_1)O_2(t_2),\ldots,O_n(t_n)} = \beta^n \left.
    \frac{\partial^n \Omega}{\partial \alpha_1\partial\alpha_2\ldots }\right|_{\alpha_i = 0}.
\end{equation}
Therefore  higher order responses of a coupled cluster approximation to $\Omega[\alpha_1, \alpha_2, \ldots]$, constructed along the appropriate contour,
can be used to obtain the coupled cluster approximation to arbitrarily general correlation functions.

\subsection{Differential form of the Keldysh CC equations}\label{sec:diff}
Thus far we have expressed everything as non-linear Volterra type integral equations, but there is an equally valid differential form (see for example the TCC representation of finite temperature coupled cluster given in references~\onlinecite{Sanyal1992,Sanyal1993,Mandal2001,Mandal2003}). We start with the integral equations for $s$ and $\tilde{\lambda}$ along the Keldysh contour:
\begin{align}
    s_{\mu}(t) &= -i\int_{C(0)}^{C(t)}dt'e^{i\Delta_{\mu}(t' - t)}\mathrm{S}_{\mu}[\bv{s}(t')] \label{eqn:Smu}\\
    \tilde{\lambda}_{\mu}(t) &= -i\int_{C(t')}^{C(-i\beta)}dt'e^{i\Delta_{\mu}(t - t')}\mathrm{L}_{\mu}[\bv{s}(t'),\tilde{\bv{\lambda}}(t')].\label{eqn:Lmu}
\end{align}
Note that since the density (Equations~\ref{eqn:g1ia}-\ref{eqn:g1ai}) depends only on $\tilde{\lambda}$, we never need to work with $\lambda$ directly using this formulation. 

Taking the time derivative of Equations~\ref{eqn:Smu} and~\ref{eqn:Lmu} yields equations of motion for $s$ and $\tilde{\lambda}$:
\begin{align}
    \frac{ds_{\mu}}{dt} &= (-i)\left\{\Delta_{\mu}s_{\mu}(t) + \mathrm{S}_{\mu}[\bv{s}(t)]\right\} \\
    \frac{d\tilde{\lambda}_{\mu}}{dt} &= i\left\{\Delta_{\mu}\tilde{\lambda}_{\mu}(t) + 
    \mathrm{L}_{\mu}[\bv{s}(t),\tilde{\bv{\lambda}}(t)] \right\}.
\end{align}
In both these equations, the term proportional to $\Delta_{\mu}$ comes from the fact that we have formulated our equations in the Schrodinger picture.

These equations can be used to directly propagate the $s$-amplitudes and $\tilde{\lambda}$-amplitudes along the Keldysh contour. In this way it is possible to avoid storing the amplitudes as is necessary when working with the integral formulation of the equations. However, working with this differential form presents other difficulties. While the value of $s_{\mu}$ is initially known to be zero at $t = 0$, the value of $\tilde{\lambda}_{\mu}$ is known at $t = -i\beta$. Considering the dependence of L$_{\mu}$ on $s$, this requires the calculation to be carried out in two steps. First $s_{\mu}$ is propagated from $t = 0$ to $t = -i\beta$. This propagation can be carried out along the Keldysh contour or just along the imaginary axis. Second, both $s$ and $\tilde{\lambda}$ are propagated back from $t = -i\beta$ to $t = 0$ along the Keldysh contour and the density matrix is computed along the way as per Appendix~\ref{sec:A1rdm}. Though this differential formulation is ultimately a more natural form for the dynamical problem, we have not yet fully investigated the numerical properties of such a method and instead use the integral formulation to investigate the basic properties and some initial applications.

\subsection{Violation of conservation laws in Keldysh-CC}\label{sec:KelConserv}
When considering dynamics, and finite-temperature dynamics in particular, it is more difficult to construct approximations that yield dynamics that match even qualitatively those of the exact system. Kadanoff and Baym\cite{Baym1961,Systems1962} stated conditions that approximate Green's functions must satisfy in order to yield dynamics consistent with all local conservation laws (such as the continuity equation relating density and current), and approximations that meet these criteria are called ``conserving''. Coupled cluster theory is  not constructed as a conserving approximation. However, coupled cluster methods have been used to obtain accurate zero-temperature dynamics in small finite systems nonetheless\cite{Schonhammer1978,Hoodbhoy1978,Hoodbhoy1979,Huber2011,Kvaal2012,Nascimento2017}. Furthermore, even conserving approximations can yield qualitatively incorrect results\cite{PuigVonFriesen2010}. We do not consider this property to be an absolute requirement for useful approximations to dynamics.

Unfortunately, the Keldysh CC presented in this work, truncated at some finite excitation level, can also violate global conservation laws. Perturbation theory approximations to the cumulant generating functional do conserve global symmetries in general as we will now show. 

Consider a cumulant generating functional, $\Omega$, that admits a perturbation theory expansion in some interaction controlled by a coupling, $\lambda$, such that
\begin{equation}
	\Omega[\lambda] = \Omega[0] + \lambda \Omega'[0] + \frac{\lambda^2}{2}\Omega''[0] + \ldots
\end{equation}
Note that $n$th order correction can be computed by finite difference in that
\begin{equation}
	\Omega^{[n]} = \left.\frac{1}{n!}\frac{\partial^n \Omega[\lambda]}{\partial\lambda^n}\right|_{\lambda = 0} = \sum_{x}c_x\Omega[\lambda_x]
\end{equation}
where the $\lambda_x$ are points slightly displaced from $\lambda = 0$, and the $\{c_x\}$, $\{\lambda_x\}$ are defined by a particular finite difference approximation. If we now compute some time-dependent observable, $O$, by differentiating with respect to a time-dependent coupling, $\alpha(t)$, the $n$th order contribution to the observable is
\begin{align}\label{eqn:final}
	\avg{O}^{[n]}(t) &= \frac{\partial}{\partial\alpha(t)} \left.\Omega^{[n]}[\alpha(t)]\right|_{\alpha(t) = 0}\nonumber \\
	&=\frac{\partial}{\partial\alpha(t)} \left.\sum_{x} c_x\Omega[\lambda_x,\alpha(t)]\right|_{\alpha(t) = 0}.
\end{align}
This implies that if $O$ is globally conserved by the exact dynamics, it should also be conserved in perturbation theory. In Appendix~\ref{sec:APconserv}, we present a detailed analysis of particle number conservation in an exactly solvable problem which suggests that in general, truncated Keldysh-CC does not have this property. This can limit the applicability of the method as we will discuss in Section~\ref{sec:apps}.

\section{Implementation}\label{sec:Implementation}
To enable a concrete implementation, we truncate the coupled cluster equations to the singles and doubles level (CCSD), where $\mu$ labels
at most four fermionic lines.
The implementation of Keldysh-CCSD closely mirrors the implementation of FT-CCSD, so we refer to Ref.~\onlinecite{White2018} for most of the details. There are two primary differences. First, everything must use complex arithmetic as a result of the real-time propagation. Second, the numerical time integration must be appropriately modified.

\subsection{Numerical integration}
Conceptually, we integrate the amplitude equations (Eq.~\ref{eqn:amplitude_eqns})
just as for FT-CCSD, except that we now consider each branch of the contour separately:
\begin{align}
    s_{\mu}^f(t_y) &= -i\sum_x G_x^ye^{i\Delta_{\mu}(t_x - t_y)}\text{S}_{\mu}[
    \bv{s}^f(t_x)] \\
    s_{\mu}^b(t_y) &= s_{\mu}^f(t_f)e^{i\Delta_{\mu}(t_f - t_y)} \nonumber \\
    &+ i\sum_{x}G_x^ye^{i\Delta_{\mu}(t_x - t_y)}\text{S}_{\mu}[\bv{s}^b(t_x)]\\
    s_{\mu}^i(\tau_y) &= s_{\mu}^b(0)e^{-\Delta_{\mu}\tau_y} \nonumber \\
    & - \sum_{x}G_x^ye^{\Delta_{\mu}(t_x - t_y)}\text{S}_{\mu}[\bv{s}^i(\tau_x)].
\end{align}
Again, we have used labels $f$, $b$, and $i$ to refer to the different branches of the Keldysh contour (forward in real time, backward in real time, and imaginary time), and as in our FT-CCSD description, we have used $G$ as a generic tensor of quadrature weights such that
\begin{equation}
    \int_{t_x}^{t_y}f(t) dt \approx \sum_{x}G_x^yf(t_x).
\end{equation}

In the non-equilibrium case we compute by differentiation the response density matrix at a particular time. This means that, unlike in the equilibrium case, we must use a quadrature rule where the weights have a smooth continuum limit,
\begin{equation}
    G_x^y \rightarrow G(t_x), \quad g_y \rightarrow g(t_y)\quad \text{as} \quad n_g \rightarrow \infty,
\end{equation}
where $G(t)$ and $g(t)$ are smooth functions.
If this condition is not satisfied, the response density matrix as a function of time will not have a well-defined limit as the grid becomes finer. This means that we cannot use quadratures like Simpson's rule or Boole's rule. In this work we use constant quadrature weights, i.e.
\begin{equation}
    g_y = 1/\delta, \qquad G_x^y = \begin{cases} 1/\delta, & x \leq y\\
    0, &\mathrm{otherwise}
    \end{cases}
\end{equation}
where $\delta$ is the grid spacing. This has the advantage of simplicity, but it requires a large number of grid points to obtain high accuracy and this limits the current study to relatively short times. 

\subsection{Computational considerations}
We will use $n_r$, $n_i$, and $n_g = 2n_r + n_i$ to indicate the number of grid points in real time, imaginary time, and in total respectively. If $n$ is the number of orbitals, the most expensive computational steps scale like $O(n_gn^6)$ and $O(n_g^2n^4)$. The $O(n_gn^6)$ step arises from the most expensive CCSD contractions which must be performed for the amplitudes at each point in time. The $(n_g^2n^4)$ step comes from the numerical integration in that the amplitudes at each point must be computed from summing quantities at all previous time points. For quadrature rules that strictly obey the integral property
\begin{equation}
    \int_0^{t_y}dt'[\ldots] = \int_0^{t_{y - 1}}dt'[\ldots] + \int_{t_{y - 1}}^{t_y}dt'[\ldots]
\end{equation}
in discretized form,
\begin{equation}
    G_x^yI_x = G_x^{y - 1}I_x + G_{y}^yI_y,
\end{equation}
the integrals to $t_{y}$ can reuse the information from the previous point and the $O(n_g^2n^4)$ step can be reduced to $O(n_gn^4)$. 

The storage requirement scales like $O(n_gn^4)$ if we store all the amplitudes as in FT-CCSD. Most of the amplitudes can be stored on disk at any given time, but the storage requirement becomes quickly prohibitive if $n_g$ is large and this limits the lengths of real time that we can consider. Using the differential form of the equations described in Section~\ref{sec:diff} can reduce the storage requirement to $O(n^4)$ and eliminate the $O(n_g^2)$ computational steps.

All computations must use complex arithmetic and, as a result, the grand potential and response properties may acquire an imaginary part. The grand potential and equilibrium properties should
be real for real-valued reference orbitals and exact integration because no operator appears on the forward and backward parts of the Keldysh contour, and any complex exponentials will cancel. 
However, this is not the case for dynamic properties,
and they can acquire an imaginary part due to errors in the coupled cluster approximation on the real axis. This is similar to the small imaginary parts of properties obtained from ground state CCSD with complex orbitals. Thus, if we use reference orbitals that are real, then we may attribute any imaginary part in the grand potential (Keldysh cumulant generating function) to  numerical integration, while for dynamic properties the imaginary part can be due to the numerical integration and to the coupled cluster approximation itself.

\section{Applications}\label{sec:apps}

\subsection{A simple benchmark}
\begin{figure}
\center
\includegraphics{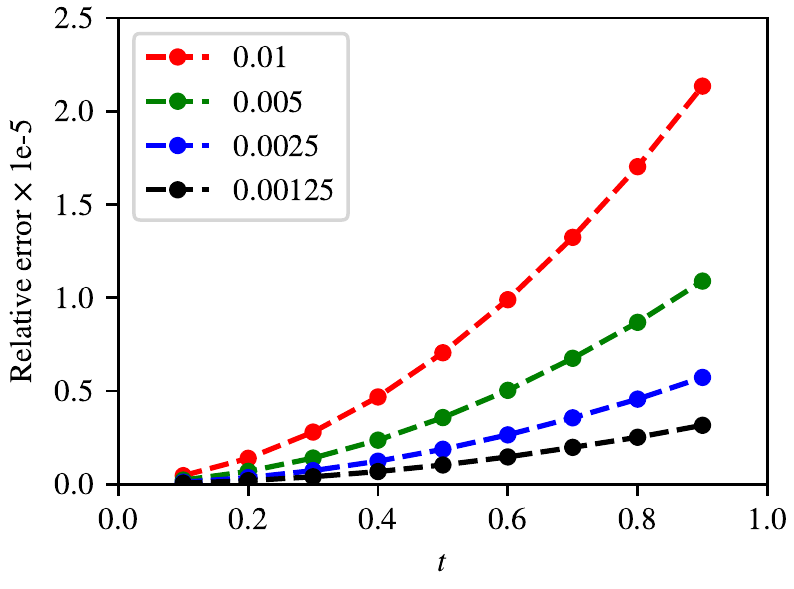}
\caption{\label{fig:error} Relative error in the dipole moment as a function of a time for a simple 2 orbital model (Equation~\ref{eqn:2orb}) for different grid spacings.}
\end{figure}

To demonstrate that the Keldysh CCSD method yields exact dynamics for a 2 orbital system, we present some benchmark calculations on a Hamiltonian of the form
\begin{align}\label{eqn:2orb}
    H(t) &= \sum_{ij} h_{ij}a^{\dagger}_ia_j + \frac{1}{2}\sum_{ijkl}
    V_{ijkl}c^{\dagger}_ia^{\dagger}_ja_la_k \nonumber \\
    &+ \sum_{ij}
    d_{ij}a^{\dagger}_ia_j \sin(\omega t)
\end{align}
where $i,j$ run over 2 spatial orbitals having the same spin degree of freedom.
We take $h$ and $V$ to be the 1-electron and 2-electron spatial-orbital integrals of a STO-3G hydrogen molecule at $R = 0.6$\AA, and $d$ to be the integrals of the dipole operator. We choose a laser frequency in the UV ($\omega = 0.2095588$) and a temperature of $k_BT = 1$. In Figure~\ref{fig:error}, we show the error in the dipole moment as a function of time relative to the full CI value obtained by explicitly propagating the full CI, grand canonical density matrix.
As expected, the theory is exact in that the only error is due to the numerical integration. An analysis of the error on a log-log plot indicates that the error is approximately quadratic with time for a constant grid spacing. 

\subsection{Hubbard model with Peierls substitution}\label{sec:Hubbard}
\begin{figure}
\center
\includegraphics{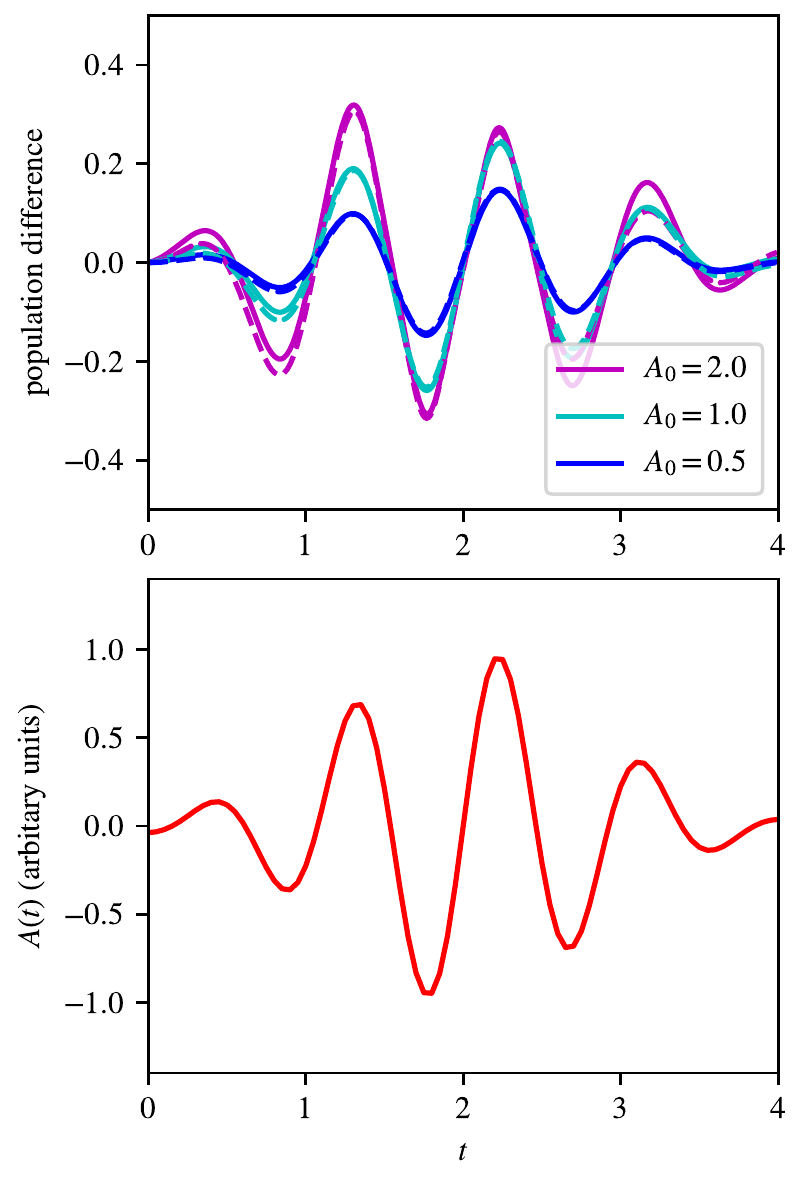}
\caption{\label{fig:HubbardExample} Population difference as a function of time for the 2-site model described in Section~\ref{sec:Hubbard}. The solid line is the exact result and the dotted lines are the Keldysh-CCSD results. Only the real part of the approximate property is plotted. In the lower panel, we show the form of the time-dependent vector potential.}
\end{figure}

The Hubbard model is one of the simplest models of correlated electrons. Despite its simplicity, the Hubbard model is valuable as a testing ground for theory and as a minimal model of correlation-induced phenomena. The interaction of the underlying lattice with a field can be
  incorporated into such a model by adding a ``Peierls phase'' so that the Hamiltonian becomes time-dependent:
\begin{equation}
    H(t) = -t_H\sum_{i\sigma}\left[ e^{iA(t)}a_{i\sigma}^{\dagger}a_{(i+1)\sigma} 
    + \text{h.c.} \right] + U\sum_{i}n_{i\uparrow}n_{i\downarrow}.
\end{equation}
The parameters $t_H$ and $U$ are the hopping and on-site repulsion respectively. We study the 1-dimensional Hubbard model with a Peierls phase of the form:
\begin{equation}
    A(t) = A_0 e^{-(t - t_0)^2/2\sigma^2}\cos[\omega(t - t_0)].
\end{equation}
This functional form is taken almost exactly from Ref.~\onlinecite{Wang2017}, where the dynamics of this model are studied at zero temperature.
We use $U = 0.5$, $\sigma = 0.8$, $t_0 = 2$, $\omega = 6.8$, and we show full CI and Keldysh-CCSD results for $A_0 = 0.5,1.0,2.0$. The simulation starts from a temperature of $T = 1.0$ at half-filling. We use zero-temperature unrestricted Hartree-Fock (UHF) orbitals and orbital energies for the coupled cluster calculation. These calculations were done with 800 grid points which is enough such that the error due to numerical integration is less than 1\%. The results for the difference in population between the two sites are shown in Figure~\ref{fig:HubbardExample}.
Even for this simple problem, Keldysh-CCSD is no longer exact, but it still performs quite well relative to the exact the result. As expected, the error is larger for larger values of $A_0$, when the system is driven further out of equilibrium. For $A_0 = 0.5$ and $A_0 = 1.0$, the deviations from the exact dynamics are barely visible in Figure~\ref{fig:HubbardExample}. Despite the fact that global symmetries are not generally conserved by Keldysh-CCSD, the particle number is conserved for this particular Hamiltonian by virtue of the particle-hole symmetry at half-filling.

\subsection{Ultrafast electron dynamics in Si at high temperature}
We now apply the method to driven electron dynamics in high-temperature silicon in order to demonstrate clearly the advantages and drawbacks of the method when applied to a more realistic Hamiltonian. We use a single primitive cell of Si in a minimal basis (SZV\cite{VandeVondele2007} with GTH pseudopotentials\cite{Goedecker1996,Hartwigsen1998}) with the ions frozen at the experimental lattice constant (3.567\AA). The matrix elements were obtained from the PySCF software package using plane-wave density fitting\cite{Sun2018}. We use the dipole approximation in the velocity gauge so that the coupling has the form:
\begin{equation}
    \frac{1}{mc}\bv{p}\cdot\bv{A}(t).
\end{equation}
$\bv{A}(t)$ is the time-dependent vector potential and it is related to the electric field by
\begin{equation}
    \bv{E}(t) = -\frac{1}{c}\dot{\bv{A}}(t).
\end{equation}

To mimic the application of an ultrafast XUV pulse, we use a pulse of the form
\begin{equation}
    \frac{1}{c}\bv{A}(t) = A_0\hat{\bv{z}}
    e^{-(t - t_0)^2/2\sigma^2}\cos[\omega(t - t_0)]
\end{equation}
with the following pulse width and center: $\sigma = 48.378$ as (2 atomic units), $t_0 = 362.83$ as (15 atomic units). $\omega$ is chosen to correspond to a 46.9 nm XUV laser. The system is taken to initially have a temperature of $T = 63155.01$ K ($k_BT = 0.2 E_{\text{h}}$). These parameters are unrealistic in two ways. We have chosen an unrealistically short laser pulse so that we can avoid doing dynamics for very long times, and we have chosen an unrealistically high temperature so that finite-temperature effects are accentuated. Despite the unphysical conditions, the Hamiltonian itself should capture the qualitative features of {\it ab initio} Hamiltonians of crystalline solids.

\begin{figure}
    \centering
    \includegraphics[scale=1.0]{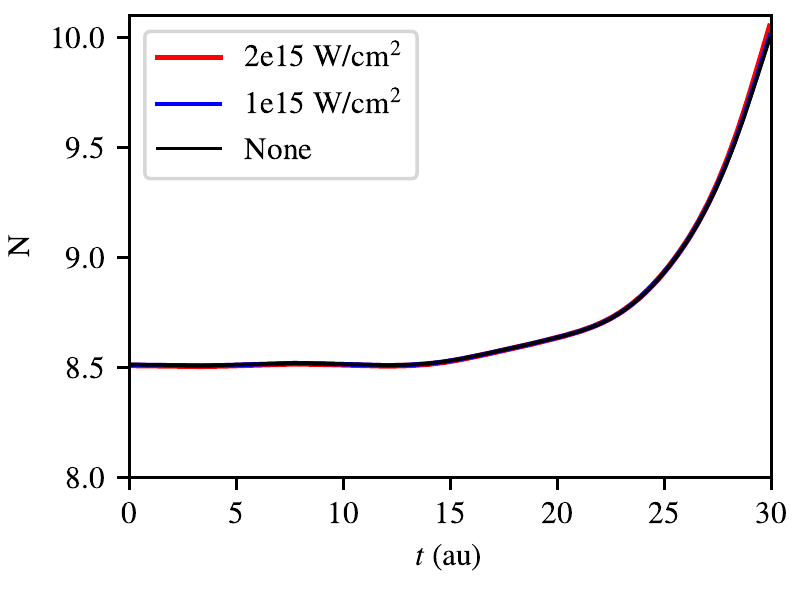}
    \caption{The number of electrons per unit cell for the Si system with different laser intensities. The anomalous drift in the electron number as a function of time is qualitatively the same for these laser intensities.}
    \label{fig:trace}
\end{figure}

We must first ask whether or not the particle number is conserved by Keldysh-CCSD dynamics, since we have already stated that this can be a problem. In Figure~\ref{fig:trace}, we have plotted the number of electrons per unit cell as a function of real-time.
Clearly the number of electrons is not conserved by Keldysh-CCSD, and the drift in the particle number is qualitatively the same for different laser intensities. We expect this to lead to unphysical results at longer times.

\begin{figure}
    \centering
    \includegraphics[scale=1.0]{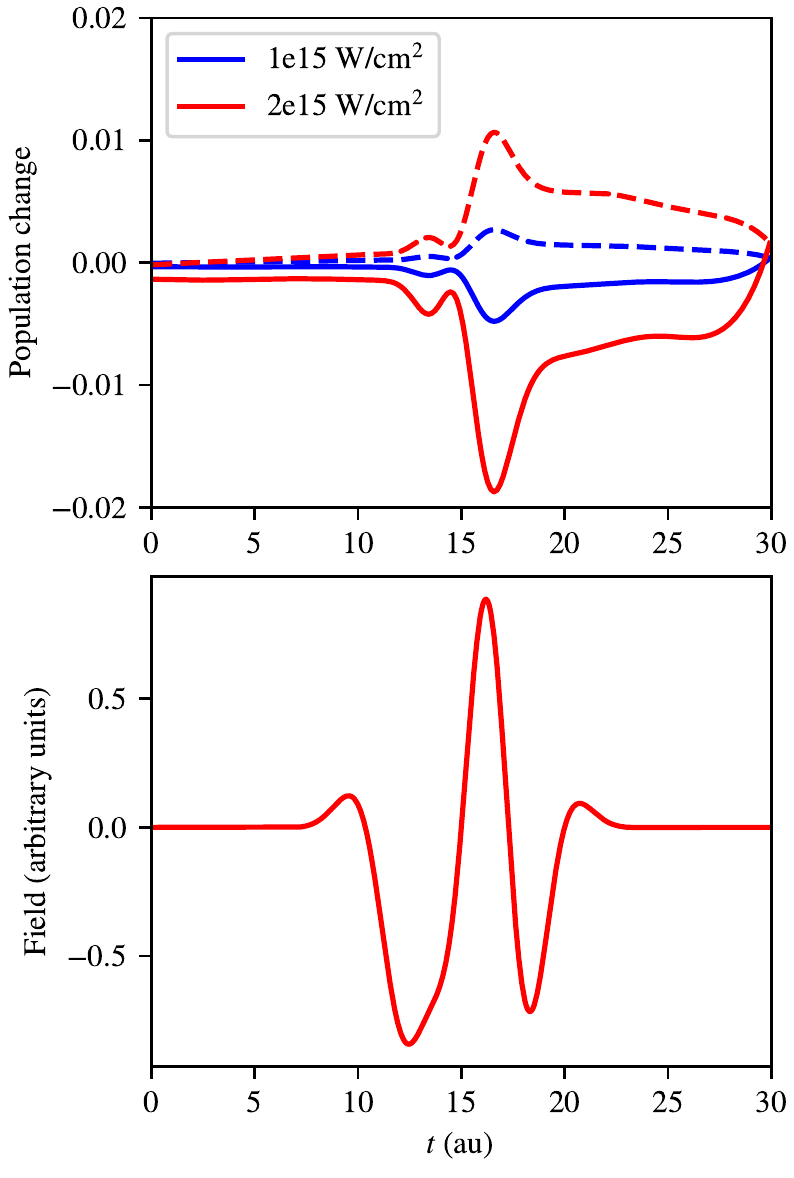}
    \caption{Difference in the population of the valence band (solid line) and conduction band (dotted line) for different maximum laser intensities as a function of time. The shape of the electric field as a function of time is also shown.}
    \label{fig:diff}
\end{figure}

If we are only interested in relatively short times, the laser-driven dynamics can be corrected by examining the difference induced by the laser field. The differences in the populations of the valence and conduction bands induced by the laser as a function of $t$ are shown in Figure~\ref{fig:diff}.
Here we see dynamics that is physically reasonable. At high intensity, the laser increases the population of the conduction band and
depopulates the valence band. Thus despite the violation of particle number conservation, Keldysh-CCSD can provide an approximation to the dynamics which yields qualitatively correct correlated many-body dynamics at finite temperature for short times. 

\section{Conclusions and future work}
In this work, we have shown how the Keldysh formalism may be used to extend imaginary time finite temperature coupled cluster to treat the finite temperature dynamics of correlated fermionic systems. We have discussed the theory in detail and presented results for the singles and doubles variant of this method, Keldysh-CCSD. In addition to the integral equations that form the basis of our current implementation, we have shown how the theory may be extended to the computation of correlation functions with arbitrary time-orderings, and we have shown how a differential form, more natural for dynamics, may be used to avoid storing the amplitudes at all times. One difficulty in applying the method to realistic systems is that there is the potential for the violation of global symmetries. In Section~\ref{sec:KelConserv} we have discussed this issue, and in Appendix~\ref{sec:APconserv}, we have presented a more detailed analysis for an exactly solvable system.

In order to demonstrate some of the potential applications and issues with the method, we have performed calculations on a 2-site Hubbard model with a Peierls phase and simple {\it ab initio} model of bulk silicon in the presence of an intense XUV laser pulse. For the small Hubbard model, Keldysh-CCSD performs well relative to the exact dynamics, while for the silicon model, we are limited to studying relatively short-time dynamics due to unphysical results at longer times.

The generalization of coupled cluster theory to finite-temperature dynamics presented here displays a number of practical and theoretical challenges which present opportunities for future work. Chief among them are
\begin{enumerate}
    \item Practical improvements necessary to decrease the computational cost and memory requirements
    \item Modifications of the theory to lessen the impact of particle-number symmetry breaking and other unphysical effects at longer times 
\end{enumerate}

\begin{acknowledgements}
This work is supported by the US Department of Energy, Office of Science, via grant number SC0018140.
\end{acknowledgements}

\appendix
\section{Introduction to the Keldysh formalism}\label{sec:AKeldysh}
We imagine a system in thermal equilibrium at time $t = 0$. At $t = 0$ some perturbation, $v(t)$, is turned on. The average value of $O$ at some time $t_f$ is given by
\begin{equation}
	\avg{O(t_f)} = \text{Tr}\left[e^{-iH_v(t_f)}\rho e^{iH_v(t_f)} O\right]
\end{equation}
where $\rho$ is the equilibrium density operator,
\begin{equation}
    \rho = e^{-\beta(H - \mu N)},
\end{equation}
for inverse temperature $\beta$ and chemical potential $\mu$, and $H_v(t)$ is the full Hamiltonian including the perturbation,
\begin{equation}
    H_v(t) \equiv H + v(t).
\end{equation}
The cyclic property of the trace allows us to write this expression as
\begin{equation}
	\avg{O(t_f)} = \text{Tr}\left[\rho O_H(t_f)\right],
\end{equation}
where the subscript $H$ indicates the Heisenberg representation.

We can now rewrite this expression in the interaction picture by splitting the Hamiltonian as
\begin{equation}
	H_v(t) = K_0 + V(t)
\end{equation}
where $K_0$ is the one-electron part,
\begin{equation}\label{eqn:K0}
    K_0 = H_0 - \mu N,
\end{equation}
and $V(t)$ is the interaction and includes $v(t)$ for $t > 0$. Note that if the Hamiltonian commutes with the number operator the chemical potential term can be included in the propagator from the beginning or trivially added at any stage. The expression for $O$ becomes
\begin{equation}
	\avg{O(t_f)} = 
	\frac{1}{Z}\text{Tr}\left[e^{-\beta K_0}
	U_I(\beta)U_I^{\dagger}(t_f)O_I(t_f)U_I(t_f)\right]
\end{equation}
where $U_I$ is the propagator in the interaction picture (as indicated by the ``I" subscript), and $Z$ is the grand canonical partition function. Inserting the time-ordered representations of the propagators, we obtain
\begin{widetext}
\begin{align}
	\avg{O(t_f)} &= \frac{1}{Z}\text{Tr}\left[e^{-\beta K_0}
	\mathcal{T}\exp\left(-\int_0^{\beta}d\tau V_I(\tau) \right)
	\mathcal{T}^{\ast}\exp\left(-i\int_{t_f}^0dt V_I(t)\right)O(t_f)
	\mathcal{T}\exp\left(-i\int_{0}^{t_f}dt V_I(t)\right)\right] \\
	&= \frac{1}{Z}\text{Tr}\left\{e^{-\beta K_0}
	\mathcal{T}_C\left[\exp\left(-i\int_C dt V_I(t)\right)O(t_f)\right]
	\right\}.\label{eqn:KelyshAvg}
\end{align}
\end{widetext}
In the first line we have used the $\mathcal{T}$ as the time-ordering operator in real or imaginary time and $\mathcal{T}^{\ast}$ as the anti-time-ordering operator. In the second line we have introduced the contour-ordering operator $\mathcal{T}_C$ which orders its arguments along the Keldysh (or Kadanoff-Baym) contour (see Figure~\ref{fig:contour}) and allows us to write the out-of-equilibrium thermal average in a compact way. The contour $C$ usually has three branches that go from $0\rightarrow t_f$, $t_f \rightarrow 0$, and $0\rightarrow -i\beta$ respectively, though many different choicwes are possible.

\section{Keldysh CCSD amplitude equations}\label{sec:aampl}
The amplitude equations are determined by the kernels S$_1$ and S$_2$, which contain contractions identical to those in the CCSD T$_1$ and T$_2$ equations:
\begin{widetext}
\begin{align}\label{eqn:ccS1}
	\mathrm{S}_i^a(t) &= f_{ai}(1 - n_a)n_i + \sum_b f_{ab}(1 - n_a)s_i^b(t) - \sum_j f_{ji}n_is_j^a(t) + 
	\sum_{jb}\bra{ja}\ket{bi}(1 - n_a)n_is_j^b(t) + \sum_{jb}f_{jb}s_{ij}^{ab}(t)\nonumber \\
	&+ \frac{1}{2}\sum_{jbc}\bra{aj}\ket{bc}(1 - n_a)s_{ij}^{bc}(t)
	- \frac{1}{2}\sum_{jkb}\bra{jk}\ket{ib}n_is_{jk}^{ab}(t)
	-\sum_{jb} f_{jb}s_i^b(t)s_j^a(t) \nonumber \\
	&+ \sum_{jbc}\bra{ja}\ket{bc}(1 - n_a)
	s_j^b(t)s_i^c(t) - \sum_{jkb}\bra{jk}\ket{bi}n_is_j^b(t)s_k^a(t) 
	-\frac{1}{2}\sum_{jkbc}\bra{jk}\ket{bc}s_i^b(t)s_{jk}^{ac}(t)\nonumber \\
	&- \frac{1}{2}\sum_{jkbc}\bra{jk}\ket{bc}s_j^a(t)s_{ik}^{bc}(t)
	+ \sum_{jkbc}\bra{jk}\ket{bc}s_j^b(t)s_{ki}^{ca}(t)
	+ \sum_{jkcd}\bra{jk}\ket{bc}s_i^b(t)s_j^c(t)s_k^a(t)
\end{align}
\begin{align}
	\mathrm{S}_{ij}^{ab}(t) &= \bra{ab}\ket{ij}(1 - n_a)(1-n_b)n_jn_j + P_{ij}\sum_{c}\bra{ab}\ket{cj}s_i^c(t)(1 - n_a)(1 - n_b)n_j \nonumber \\
	&- P_{ab}\sum_{k} \bra{kb}\ket{ij}(1 - n_b)n_in_js_k^a(t) + P_{ab}\sum_{c}f_{bc}(1 - n_b)s_{ij}^{ac}(t) \nonumber \\
	&- P_{ij}\sum_{k}f_{kj}n_js_{ik}^{ab}(t) + 
	\frac{1}{2}\sum_{cd}\bra{ab}\ket{cd}(1 - n_a)(1 - n_b)s_{ij}^{cd}(t) \nonumber \\
	&+\frac{1}{2}\sum_{kl}\bra{kl}\ket{ij}n_in_js_{kl}^{ab}(t)
	+ P_{ij}P_{ab}\sum_{kc}\bra{kb}\ket{cj}(1 - n_b)n_js_{ik}^{ac}(t)\nonumber \\
	&+ \frac{1}{2}P_{ij}\sum_{cd}\bra{ab}\ket{cd}(1 - n_a)(1 - n_b)s_i^c(t)s_j^d(t)
	+ \frac{1}{2}P_{ab}\sum_{kl}\bra{kl}\ket{ij}n_in_js_k^a(t)s_l^b(t) 
	\nonumber\\ 
	&- P_{ij}P_{ab}\sum_{kc}\bra{ak}\ket{cj}(1 - n_a)n_js_i^c(t)s_k^b(t)
	-P_{ij}\sum_{kc}f_{kc}s_i^c(t)s_{kj}^{ab}(t)
	- P_{ab}\sum_{kc}f_{kc}s_k^as_{ij}^{cb}(t)\nonumber \\
	&+ P_{ab}\sum_{kcd}\bra{ka}\ket{cd}(1 - n_a)s_k^c(t)s_{ij}^{db}(t)
	- P_{ij}\sum_{klc}\bra{kl}\ket{ci}n_is_k^c(t)s_{lj}^{ab}(t) 
	+ P_{ij}P_{ab}\sum_{kcd}\bra{ak}\ket{cd}(1 - n_a)s_i^c(t)s_{kj}^{db}(t) \nonumber \\
	&- P_{ij}P_{ab}\sum_{klc}\bra{kl}\ket{ic}n_is_k^a(t)s_{lj}^{cb}(t) 
	+ \frac{1}{2}P_{ij}\sum_{klc}\bra{kl}\ket{cj}n_js_i^c(t)s_{kl}^{ab}(t)
	- \frac{1}{2}P_{ab}\sum_{kcd}\bra{kb}\ket{cd}(1 - n_b)s_k^a(t)s_{ij}^{cd}(t)\nonumber \\
	&+ \frac{1}{4}\sum_{klcd}\bra{kl}\ket{cd}s_{ij}^{cd}(t)s_{kl}^{ab}(t)
	+ \frac{1}{2}P_{ij}P_{ab}\sum_{klcd}\bra{kl}\ket{cd}s_{ik}^{ac}(t)s_{lj}^{db}(t)
	- \frac{1}{2}P_{ab}\sum_{klcd}\bra{kl}\ket{cd}s_{kl}^{ca}(t)s_{ij}^{db}(t)\nonumber \\ 
	&- \frac{1}{2}P_{ij}\sum_{klcd}\bra{kl}\ket{cd}s_{ki}^{cd}(t)s_{lj}^{ab}(t)
	-\frac{1}{2}P_{ij}P_{ab}\sum_{kcd}\bra{kb}\ket{cd}(1 - n_b)s_i^c(t)s_k^a(t)s_j^d(t)
	+ \frac{1}{2}P_{ij}P_{ab}\sum_{klc}\bra{kl}\ket{cj}n_js_i^c(t)s_k^a(t)s_l^b(t) \nonumber\\
	&+\frac{1}{4}P_{ij}\sum_{klcd}\bra{kl}\ket{cd}s_i^c(t)s_j^d(t)s_{kl}^{ab}(t)
	+ \frac{1}{4}P_{ab}\sum_{klcd}\bra{kl}\ket{cd}s_k^a(t)s_l^b(t)s_{ij}^{cd}(t)
	-P_{ij}P_{ab}\sum_{klcd}\bra{kl}\ket{cd}s_i^c(t)s_k^a(t)s_{lj}^{db}(t)	\nonumber \\
	&- P_{ij}\sum_{klcd}\bra{kl}\ket{cd}s_k^c(t)s_i^d(t) s_{lj}^{ab}(t)
	- P_{ab}\sum_{klcd}\bra{kl}\ket{cd}s_k^c(t)s_l^a(t)s_{ij}^{db}(t)
	+\frac{1}{4}P_{ij}P_{ab}\sum_{klcd}\bra{kl}\ket{cd}
	s_i^c(t)s_k^a(t)s_l^b(t)s_j^d(t).
\end{align}
\end{widetext}
We have used $n_p$ to indicate the Fermi-Dirac occupation number of the $p$th orbital:
\begin{equation}
    n_p \equiv \frac{1}{e^{\beta(\varepsilon_p - \mu)} + 1}.
\end{equation}
Here and elsewhere, we use the notation:
\begin{equation}
    P_{ab}F[ab] = F[ab] - F[ba].
\end{equation}

\section{Keldysh CCSD lambda equations}\label{sec:Alambda}
In practice, we solve for the discretized Lagrange multipliers, $\lambda(t_y)$, by taking derivatives of the discretized Lagrangian with respect to the amplitudes at each grid point. This leads to equations of the form
\begin{align}
    \lambda^f_{\mu}(t_x) &= -\text{L}_{\mu}[\bv{s}_f(t_x),\tilde{\lambda}_f(t_x)] \\
    \lambda^b_{\mu}(t_x) &= -\text{L}_{\mu}[\bv{s}_b(t_x),\tilde{\lambda}_b(t_x)] \\
    \lambda^i_{\mu}(\tau_x) &= -\text{L}_{\mu}[\bv{s}_i(\tau_x),\tilde{\lambda}_i(\tau_x)]
\end{align}
where we have again used $f,b,i$ to indicate forward, backward, and imaginary time branches. The $L^\mu$ kernel is local in time and contains precisely those contractions that appear in the ground state CC $\lambda$ equations with the slight modification that those terms arising from the derivative of the energy are multiplied by a negative sign:
\begin{widetext}
\begingroup
\allowdisplaybreaks
\begin{align}\label{eqn:ccL1}
	L_a^i(t) &= -f_{ia} + \sum_b \tilde{\lambda}^i_b(t)(1 - n_b)f_{ba} - \sum_j \tilde{\lambda}^j_a(t)n_jf_{ij}
	+ \sum_{jb}\tilde{\lambda}^j_b(t)(1 - n_b)n_j\bra{bi}\ket{ja} - \sum_{jb}\bra{ij}\ket{ab}s_j^b(t)
	\nonumber \\
	&- \sum_{jb}\tilde{\lambda}^j_a(t) f_{ib}s_j^b(t) - \sum_{jkb}
	\tilde{\lambda}^i_b(t) f_{ja}s_j^b(t) + \sum_{jbc}\tilde{\lambda}^i_c(t)(1 - n_c)\bra{cj}\ket{ab}s_j^b(t)
	 -\sum_{jkb}\tilde{\lambda}^k_a(t)n_k\bra{ij}\ket{kb}s_j^b(t)\nonumber \\ 
	 &+ \sum_{jbc}
	\tilde{\lambda}^j_c(t)(1 - n_c)\bra{ci}\ket{ba}s_j^b(t) - \sum_{jkb}\tilde{\lambda}^k_b(t)n_k\bra{ji}
	\ket{ka}s_j^b(t)- \frac{1}{2}\sum_{jkbc} \tilde{\lambda}^j_a(t)n_k\bra{ik}
	\ket{bc}s_{jk}^{bc}(t) \nonumber \\
	&- \frac{1}{2}\sum_{jkbc}\tilde{\lambda}^i_b(t)\bra{jk}\ket{ac}
	s_{jk}^{bc}(t) + \sum_{jkbc}\tilde{\lambda}^j_b(t)\bra{ki}\ket{ca}s_{jk}^{bc}(t)
	- \sum_{jkbc} \tilde{\lambda}^j_a(t)\bra{ik}
	\ket{bc}s_j^b(t)s_k^c(t) \nonumber \\
	&- \sum_{jkbc}\tilde{\lambda}^i_b(t)\bra{jk}\ket{ac}
	s_j^b(t)s_k^c(t) - \sum_{jkbc}\tilde{\lambda}^k_b(t)\bra{ji}\ket{ca}s_j^b(t)s_k^c(t)
	+ \frac{1}{2}\sum_{jbc}\tilde{\lambda}_{cb}^{ij}(t)(1 - n_b)(1 - n_c)n_j\bra{cb}
	\ket{aj} \nonumber \\
	&- \frac{1}{2}\sum_{jkb}\tilde{\lambda}_{ab}^{kj}(t)(1 - n_b)n_jn_k\bra{ib}\ket{kj}
	- \sum_{jkbc}\tilde{\lambda}^{jk}_{ac}(t)(1 - n_c)n_k\bra{ic}\ket{bk}
	s_j^b(t) - \sum_{jkbc}\tilde{\lambda}^{ik}_{bc}(t)(1 - n_c)n_k\bra{jc}\ket{ak}s_j^b(t) \nonumber \\
	&+ \frac{1}{2}\sum_{jbcd}\tilde{\lambda}^{ij}_{cd}(t)(1 - n_b)(1 - n_c)\bra{cd}\ket{ab}s_j^b(t)
	+ \frac{1}{2}\sum_{jklb}\tilde{\lambda}^{kl}_{ab}(t)n_kn_l\bra{ij}\ket{kl}s_j^b(t)
	-\frac{1}{2}\sum_{jkbc}\tilde{\lambda}^{jk}_{ba}(t)f_{ic}s_{jk}^{bc}(t)\nonumber \\
	&- \frac{1}{2}\sum_{jkbc}\tilde{\lambda}^{ji}_{bc}(t)f_{ka}s_{jk}^{bc}(t)
	+\frac{1}{2}\sum_{jkbcd}\tilde{\lambda}^{jk}_{bd}(t)(1 - n_d)\bra{di}\ket{ca}s_{jk}^{bc}(t)
	-\frac{1}{2}\sum_{jklbc}\tilde{\lambda}^{jl}_{bc}(t)n_l\bra{ki}\ket{la}s_{jk}^{bc}(t)
	\nonumber \\
	&+ \sum_{jkbcd}\tilde{\lambda}^{ji}_{bd}(t)(1 - n_d)\bra{kd}\ket{ca}s_{jk}^{bc}(t)
	- \sum_{jklbc}\tilde{\lambda}^{jl}_{ba}(t)n_l\bra{ki}\ket{cl}s_{jk}^{bc}(t)
	- \frac{1}{4}\sum_{jkbcd}\tilde{\lambda}^{jk}_{ad}(t)(1 - n_d)\bra{id}\ket{bc}t_{jk}^{bc}(t)
	\nonumber \\ 
	&+\frac{1}{4}\sum_{jklbc}\tilde{\lambda}^{il}_{bc}(t)n_l\bra{jk}\ket{al}s_{jk}^{bc}(t)
	-\sum_{jkbcd}\tilde{\lambda}^{ik}_{db}(t)(1 - n_d)\bra{dj}\ket{ac}s_j^b(t)s_k^c(t)
	+ \sum_{jklbc}\tilde{\lambda}^{lk}_{ab}(t)n_l\bra{ij}\ket{lc}s_j^b(t)s_k^c(t)\nonumber \\
	&- \frac{1}{2}\sum_{jkbcd}\tilde{\lambda}^{jk}_{ad}(t)(1 - n_d)\bra{id}\ket{bc}s_j^bs_k^c(t)
	+ \frac{1}{2}\sum_{jklbc}\tilde{\lambda}^{il}_{bc}(t)\bra{jk}\ket{ad}s_j^b(t)s_k^c(t)
	-\frac{1}{2}\sum_{jklbcd}
	\tilde{\lambda}^{kl}_{ca}(t)\bra{ij}\ket{db}s_j^b(t)s_{kl}^{cd}(t)\nonumber \\
	&- \frac{1}{2}\sum_{jklbcd}\tilde{\lambda}^{ki}_{cd}(t)\bra{lj}\ket{ab}s_j^b(t)s_{kl}^{cd}(t)
	- \sum_{jklbcd}\tilde{\lambda}^{jl}_{ad}(t)\bra{ik}\ket{bc}s_j^b(t)s_{kl}^{cd}(t)
	-\sum_{jklbcd}\tilde{\lambda}^{il}_{bd}(t)\bra{jk}\ket{ac}s_j^b(t)s_{kl}^{cd}(t)\nonumber \\
	&+\frac{1}{4}\sum_{jklbcd}\tilde{\lambda}^{kl}_{ab}(t)n_k\bra{ij}\ket{cd}s_j^b(t)s_{kl}^{cd}(t)
	+\frac{1}{4}\sum_{jklbcd}\tilde{\lambda}^{ij}_{cd}(t)\bra{kl}\ket{ab}s_j^b(t)s_{kl}^{bc}(t)
	-\frac{1}{2}\sum_{jklbcd}\tilde{\lambda}^{kl}_{cb}(t)\bra{ji}\ket{da}s_j^b(t)s_{kl}^{cd}(t)
	\nonumber \\
	&-\frac{1}{2}\sum_{jklbcd}\tilde{\lambda}^{kj}_{cd}(t)\bra{li}\ket{ba}s_j^b(t)s_{kl}^{cd}(t)
	+ \frac{1}{2}\sum_{jklbcd}\tilde{\lambda}^{jl}_{ac}(t)
	\bra{ik}\ket{bd}s_j^b(t)s_k^c(t)s_l^d(t) + \frac{1}{2}\sum_{jklbcd}\tilde{\lambda}^{il}_{bc}(t)
	\bra{jk}\ket{ad}s_j^b(t)s_k^c(t)s_l^d(t)
\end{align}
\endgroup
\begin{align}\label{eqn:ccL2}
	L_{ab}^{ij}(t) &= -\bra{ij}\ket{ab} + P_{ij}P_{ab}f_{ia}\tilde{\lambda}^j_b(t) + 
	P_{ij}\sum_c \tilde{\lambda}^i_c(t)(1 - n_c)\bra{cj}\ket{ab} - P_{ab}\sum_k\tilde{\lambda}^k_a(t)n_k\bra{ij}
	\ket{kb} \nonumber \\
	&+ P_{ij}P_{ab}\sum_{kc}\tilde{\lambda}^j_b(t)\bra{ik}\ket{ac}
	s_k^c(t) - P_{ij}\sum_{kc}\tilde{\lambda}^i_c(t)\bra{kl}\ket{ab}s_k^c(t) - P_{ab}\sum_{kc}
	\tilde{\lambda}^k_a\bra{ij}\ket{cb}s_k^c(t) \nonumber \\
	&+ P_{ab}\sum_c \tilde{\lambda}^{ij}_{ac}(t)(1 - n_c)f_{cb} 
	- P_{ij}\sum_k\tilde{\lambda}^{ik}_{ab}(t)n_kf_{jk} 
	+ \frac{1}{2}\sum_{cd} \tilde{\lambda}^{ij}_{cd}(t)(1 - n_c)(1 - n_d) \bra{cd}\ket{ab} + 
	\frac{1}{2}\sum_{kl}\tilde{\lambda}^{kl}_{ab}(t)n_kn_l\bra{ij}\ket{kl} \nonumber \\
	&+ P_{ij}P_{ab}\sum_{kc}\tilde{\lambda}^{ik}_{ac}(t)(1 - n_c)n_k\bra{cj}\ket{kb}
	-P_{ij}\sum_{kc}\tilde{\lambda}^{ik}_{ab}(t)f_{jc}s_k^c(t)
	- P_{ab}\sum_{kc}\tilde{\lambda}^{ij}_{ac}(t)f_{kb}s_k^c(t)\nonumber \\
	&+ P_{ab}\sum_{kcd}\tilde{\lambda}^{ij}_{ad}(t)(1 - n_d)\bra{dk}\ket{bc}s_k^c(t) - 
	P_{ij}\sum_{klc}\tilde{\lambda}^{il}_{ab}(t)n_l\bra{jk}\ket{lc}s_k^c(t)
	+ P_{ij}P_{ab}\sum_{kcd}\tilde{\lambda}^{ik}_{ad}(t)(1 - n_d)\bra{dj}\ket{cb}s_k^c(t) \nonumber \\
	&- P_{ij}P_{ab}\sum_{klc}\tilde{\lambda}^{il}_{ac}(t)n_l\bra{kj}\ket{lb}s_k^c(t) 
	- \sum_{kcd}\tilde{\lambda}^{ij}_{cd}(t)(1 - n_d)\bra{kd}\ket{ab}s_k^c(t)
	+ \sum_{klc}\tilde{\lambda}^{kl}_{ab}(t)\bra{ij}\ket{cd}s_k^c(t) \nonumber \\
	&- P_{ij}\frac{1}{2}\sum_{klcd}\tilde{\lambda}^{ik}_{ab}(t)
	\bra{jl}\ket{cd}s_{kl}^{cd} (t)- P_{ab}\frac{1}{2}\sum_{klcd}\tilde{\lambda}^{ij}_{ac}(t)
	\bra{kl}\ket{bd}s_{kl}^{cd}(t) + P_{ij}P_{ab}\sum_{klcd}\tilde{\lambda}^{ik}_{ac}(t)
	\bra{lj}\ket{db}s_{kl}^{cd}(t)\nonumber \\
	&- P_{ab}\frac{1}{2}\sum_{klcd}\tilde{\lambda}^{kl}_{ca}(t)\bra{ij}\ket{db}s_{kl}^{cd}(t)
	- P_{ij}\frac{1}{2}\sum_{klcd}\tilde{\lambda}^{ki}_{cd}(t)\bra{lj}\ket{ab}s_{kl}^{cd}(t)
	+ \frac{1}{4}\sum_{klcd}\tilde{\lambda}^{kl}_{ab}(t)\bra{ij}\ket{cd}s_{kl}^{cd}(t)
	\nonumber \\
	&+ \frac{1}{4}\sum_{klcd}\tilde{\lambda}^{ij}_{cd}(t)\bra{kl}\ket{ab}s_{kl}^{cd}(t)
	-P_{ij}\sum_{klcd}\tilde{\lambda}^{ik}_{ab}(t)
	\bra{jl}\ket{cd}s_k^c(t)s_l^d(t) - P_{ab}\sum_{klcd}\tilde{\lambda}^{ij}_{ac}(t)\bra{kl}
	\ket{bd}s_k^c(t)s_l^d(t) \nonumber \\
	&- \sum_{klcd}\tilde{\lambda}^{ik}_{ad}(t)\bra{lj}\ket{cb}s_k^c(t)s_l^d(t)
	+ \frac{1}{2}\sum_{klcd}\tilde{\lambda}^{kl}_{ab}(t)\bra{ij}\ket{cd}s_k^c(t)s_l^d(t)
	+ \frac{1}{2}\sum_{klcd}\tilde{\lambda}^{ij}_{cd}(t)\bra{kl}\ket{ab}s_k^c(t)s_l^d(t)
\end{align}
\end{widetext}
The $\tilde{\lambda}$ are integrals of the $\lambda$ amplitudes with an exponential factor. We compute them separately along each branch of the contour:
\begin{align}
    \tilde{\lambda}^f_{\mu}(t_x) &=  \sum_yg_y\lambda^i_{\mu}(\tau_y)\frac{G_x^{t_f}}{g_x}
    e^{\Delta_{\mu}(it_x - \tau_y)}\nonumber \\
    & - i\sum_yg_y\lambda^b_{\mu}(t_y)\frac{G_x^{t_0}}{g_x}
    e^{i\Delta_{\mu}(t_x - t_y)}\nonumber \\
    &+ i\sum_y g_y \lambda^f_{\mu}(t_y)\frac{G_x^y}{g_x}e^{i\Delta_{\mu}(t_x - t_y)}\\
    \tilde{\lambda}^b_{\mu}(t_x) &= \sum_yg_y\lambda^i_{\mu}(\tau_y)\frac{G_x^{t_f}}{g_x}
    e^{\Delta_{\mu}(it_x - \tau_y)} \nonumber \\
    & -i\sum_y g_y \lambda^b_{\mu}(t_y)\frac{G_x^y}{g_x}e^{i\Delta_{\mu}(t_x - t_y)}\\
    \tilde{\lambda}^i_{\mu}(\tau_x) &= \sum_y g_y \lambda^i_{\mu}(\tau_y)\frac{G_x^y}{g_x}e^{\Delta_{\mu}(\tau_x - \tau_y)}
\end{align}
The $t_f$ and $t_0$ appearing as indices of grid point tensors are those grid points occuring at $t = t_f$ and $t = 0$ respectively.

\section{Keldysh CCSD 1-RDM}\label{sec:A1rdm}
Given the $s_\mu$ and $\lambda_\mu$ amplitudes on the Keldysh contour, we can efficiently form the one-electron density matrix as a function of real time so that any one-electron observable can be computed at any real time. For Keldysh-CCSD, the coupled cluster contribution to the density has 4 parts:
\begin{widetext}
\begin{align}
	\gamma_{ia}(t_y) &= -\tilde{\lambda}^i_a(t_y) \label{eqn:g1ia}\\
	\gamma_{ba}(t_y) &= -\tilde{\lambda}^i_a(t_y)s_i^b(t_y) - \frac{1}{2}
	\tilde{\lambda}^{ki}_{cb}(t_y)s^{ca}_{ki}(t_y) \\
	\gamma_{ji}(t_y) &= \tilde{\lambda}_a^j(t_y)s_i^a(t_y) + \frac{1}{2}
	\tilde{\lambda}^{kj}_{ca}(t_y)s_{ki}^{ca}(t_y)\\
	\gamma_{ai}(t_y) &= s_i^a(t_y) - \tilde{\lambda}_b^j(t_y)s_{ji}^{ba}(t_y) +\tilde{\lambda}^j_b(t_y)s_i^b(t_y)s_j^a(t_y)\nonumber \\
	&+ \frac{1}{2}\tilde{\lambda}^{jk}_{bc}(t_y)
	s_i^b(t_y)s_{jk}^{ac}(t_y)
	+ \frac{1}{2}\tilde{\lambda}^{jk}_{bc}(t_y)
	s_j^a(t_y)s_{ik}^{bc}(t_y)\label{eqn:g1ai}
\end{align}
The value of some one-electron observable at time $t_y$ can be computed as
\begin{equation}
    \avg{O(t_y)} = \avg{O}_0 
    + \sum_{ia}\gamma_{ia}(t_y)O_{ai}n_i(1 - n_a) 
    + \sum_{ba}\gamma_{ba}(t_y)O_{ab}(1 - n_a)
    + \sum_{ij}\gamma_{ji}(t_y)O_{ij}n_j
    + \sum_{ai}\gamma_{ai}(t_y)O_{ia}
\end{equation}
\end{widetext}
We have used $\avg{\ldots}_0$ as the average in the thermal reference. Note that the density matrix as a function of real time can be computed from either the forward or backward parts of the contour. In the limit of exact integration the result will be the same, but in practice there could be differences. Averaging the values obtained from the forward and backward parts of the contour may provide slightly higher accuracy at the cost of computing additional density matrices. We have not explored this possibility and compute the densities from the forward branch of the contour.

\section{An analysis of particle number conservation in a 1-particle problem}
\label{sec:APconserv}
In this appendix we will show some illuminating analytic results for a 1-particle Hamiltonian. In particular, we will show explicitly that perturbation theories truncated at 2nd order and at 3rd order preserve the particle number, and, though Keldysh coupled cluster singles (CCS) is exact, linearized Keldysh coupled cluster singles (LCCS) is not exact and does not preserve particle number at finite temperature. These results are supported by numerical calculations, and we use the features of this simple problem to infer some properties of more general Keldysh CC models.

\subsection{Preliminaries}
We will consider a 1-particle Hamiltonian of the form
\begin{equation}
	H = h_0 + f(t).
\end{equation}
The corresponding grand potential is denoted by $\Omega$, and 
and we will can do perturbation theory in powers of $f$:
\begin{equation}
	\Omega = \Omega^{[0]} + \Omega^{[1]} + \Omega^{[2]} + \ldots
\end{equation}
The value of some observable at time $t$ is given by
\begin{equation}
	\avg{O}(t_1) = \left.\frac{\partial}{\partial \alpha} \Omega[\alpha]\right|_{\alpha = 0}
\end{equation}
where
\begin{equation}
	H[\alpha] = h_0 + f(t) + \alpha \delta_C(t - t_1)O
\end{equation}
as discussed in Section~\ref{sec:Keldysh_CC}. For the purpose of this discussion, we will assume the perturbation, $f$, is independent of time. This will simplify the algebra and make violations of particle-number conservation more obvious.

\begin{figure}
	\includegraphics[scale=0.7]{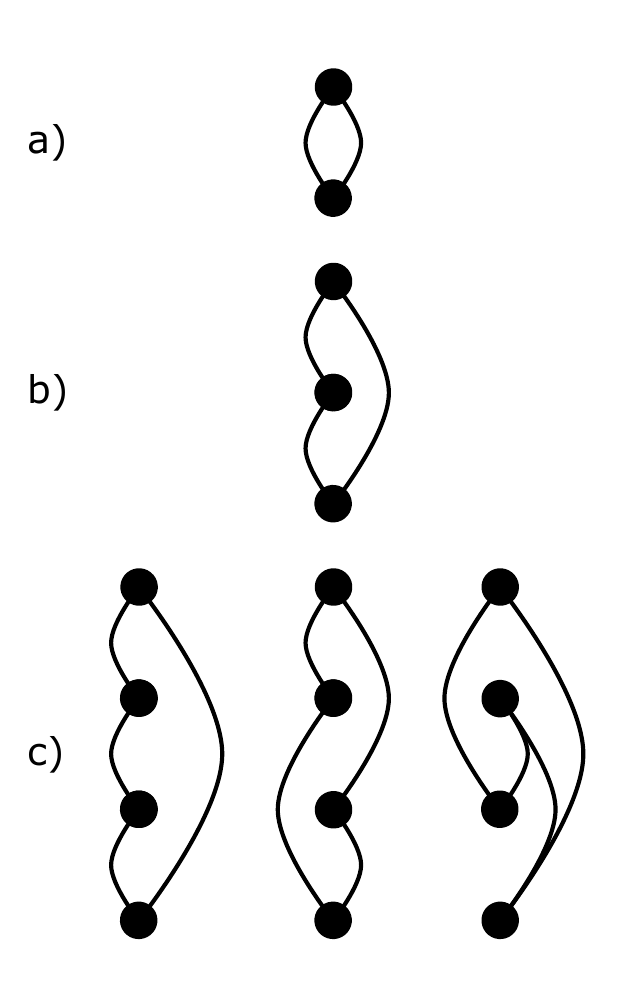}
	\caption{Hugenholtz skeletons contributing to (a) second order, (b) 3rd order, and (c) 4th order in perturbation theory.}\label{fig:pt}
\end{figure}

We will consider several theories based on perturbation theory and coupled cluster expansions. We will not derive these theories in detail, as they can be found elsewhere (see for example Ref.~\onlinecite{White2018}), but they are most easily considered from a diagrammatic perspective. The 0th and 1st order perturbation theory contributions are trivial, and are not considered here. The diagrams representing 2nd, 3rd, and 4th order perturbation theory contributions are shown in Figure~\ref{fig:pt}.  

\begin{figure}
	\includegraphics[scale=0.7]{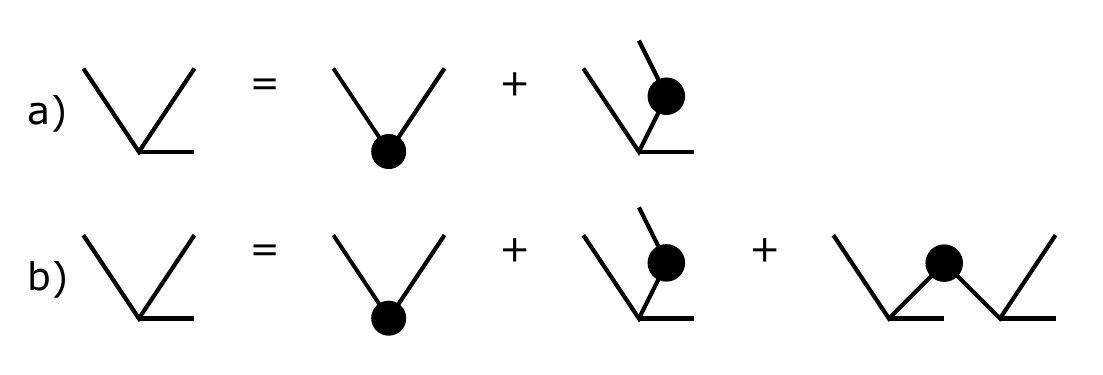}
	\caption{Hugenholtz skeletons representing the iteration in (a) LCCS and (b) CCS.}\label{fig:cc}
\end{figure}

The CCS and LCCS equations are shown diagrammatically in Figure~\ref{fig:cc}. 
A cursory analysis of these diagrams indicates that
\begin{enumerate}
\item CCS is complete to all orders in perturbation theory and is therefore exact
\item LCCS is complete only to 3rd order in perturbation theory
\item The 4th order terms missing in LCCS are those arising from the 3rd skeleton in Figure~\ref{fig:pt}(c).
\end{enumerate}

\subsection{Second order perturbation theory}
The free energy at second order can be expressed as a contour integral
\begin{equation}
	\Omega^{[2]} = \frac{i}{\beta}\int_C dt \sum_{ia} f_{ia} s_i^a(t)^{[1]} 
\end{equation}
where $s_i^a(t)^{[1]}$ are the first order amplitudes, and $C$ is a contour integral along the Keldysh contour. If we allow the perturbation to become a function of a coupling to the number operator acting at a particular point on the Keldysh contour, we get that
\begin{align}
	f_{pq}(\alpha) &= f_{pq} + i\alpha \delta_C(t - t_1)\delta_{pq} 
	\nonumber\\
	&\Rightarrow \quad \Omega^{[2]}[\alpha] 
	= \frac{i}{\beta}\int_C dt \sum_{ia} f_{ia}(\alpha) s_i^a(t)^{[1]} 
\end{align}
where the amplitudes are now implicitly a function of $\alpha$ as well. The amplitudes are equal to 
\begin{equation}\label{eqn:s1}
	s_i^a(t)^{[1]} = -i\int_{C(0)}^{C(t)}dt'e^{i\Delta_i^a(t' - t)}f_{ai}(\alpha)n_i(1 - n_a)
\end{equation}
which implies that the free energy as a function of $\alpha$ is
\begin{align}
	\Omega^{[2]}(\alpha) 
	&= \frac{1}{\beta}\int_C dt \sum_{ia} f_{ia}(\alpha)\nonumber \\
	&\times\int_{C(0)}^{C(t)}dt'
	e^{i\Delta_i^a(t' - t)}f_{ai}(\alpha)n_i(1 - n_a).
\end{align}
Expanding $\Omega$ in $\alpha$ results in
\begin{equation}
	\Omega^{[2]}(\alpha) = \Omega^{[2]}_0 + \Omega^{[2]}_1\alpha + \frac{1}{2}\Omega^{[2]}_2 \alpha^2 + \ldots 
\end{equation}
where $\Omega^{[2]}_n$ is the $n$th derivative wrt $\alpha$. We find two contributions to the first derivative:
\begin{widetext}
\begin{equation}
	\Omega_1^{[2]} = \frac{1}{\beta}\sum_{ia}n_i(1 - n_a)\left[
	\int_C dt f_{ia}\int_{C(0)}^{C(t)}dt'
	e^{i\Delta_i^a(t' - t)}i\delta_C(t' - t_1)\delta_{pq} + 
	\int_C dt i\delta_C(t - t_1)\delta_{pq}\int_{C(0)}^{C(t)}dt'
	e^{i\Delta_i^a(t' - t)}f_{ai}\right].
\end{equation}
The sum over $i,a$ reduces to a single sum because of the delta functions and we obtain
\begin{align}
	\Omega_1^{[2]} &= \frac{i}{\beta}\sum_{p}n_p(1 - n_p)\left[
	\int_C dt f_{pp}\int_{C(0)}^{C(t)}dt'
	\delta_C(t' - t_1) + 
	\int_C dt \delta_C(t - t_1)\int_{C(0)}^{C(t)}dt'
	f_{pp}\right] \\
	&= \frac{i}{\beta}\sum_{p}n_p(1 - n_p)f_{pp}\left[
	\int_{C(t_1)}^{C(-i\beta)} dt  + \int_{C(0)}^{C(t_1)}dt'\right] \\
	&= \frac{i}{\beta}\sum_{p}n_p(1 - n_p)f_{pp}\left[
	\int_Cdt\right] = \sum_{p}n_p(1 - n_p)f_{pp}.
\end{align}
\end{widetext}

This derivative is the 2nd order correction to the particle number in that
\begin{equation}
	\avg{N}_0 - \beta\Omega_1^{[2]} = 
	\avg{N}_0 - \beta\sum_pn_p(1 - n_p)f_{pp} 
\end{equation}
Notice that this correction is non-zero at finite temperature, equal to the FT-MP2 correction to the particle number (cf. Equation 49 of Ref.~\onlinecite{Santra2017}), but independent of time ($t_1$).

\subsection{Third order perturbation theory}\label{sec:mp3}
We can perform the same analysis for the 3rd order terms. We must use the second order amplitudes
\begin{widetext}
\begin{align}
	s_i^a(t)^{[2]} &= -i\int_{C(0)}^{C(t)} dt' e^{i\Delta_i^a(t' - t)}\left[\sum_b f_{ab}(\alpha)s_i^b(t')^{[1]} - \sum_j f_{ji}(\alpha)s_j^a(t')^{[1]}\right]\\
	&= -\int_{C(0)}^{C(t)} dt'e^{i\Delta_i^a(t' - t)} \Bigg[\sum_b (1 - n_a)f_{ab}(\alpha)\int_{C(0)}^{C(t')}dt''e^{i\Delta_i^b(t'' - t')}f_{bi}(\alpha)n_i(1 - n_b) \nonumber\\
	&\hspace{66pt} - \sum_j n_if_{ji}(\alpha)\int_{C(0)}^{C(t')}dt''e^{i\Delta_j^a(t'' - t')}f_{aj}(\alpha)n_j(1 - n_a)\Bigg].
\end{align}
The corresponding contribution to the grand potential is
\begin{equation*}
	\Omega^{[3]} = -\frac{i}{\beta}\int_C dt \sum_{ia} f_{ia}(\alpha)
	\int_{C(0)}^{C(t)} dt'e^{i\Delta_i^a(t' - t)}\sum_b (1 - n_a)f_{ab}(\alpha)
	\int_{C(0)}^{C(t')}dt''e^{i\Delta_i^b(t'' - t')}f_{bi}(\alpha)n_i(1 - n_b)
\end{equation*}
\begin{equation}
	\hspace{22pt} + \frac{i}{\beta}\int_C dt \sum_{ia} f_{ia}(\alpha)
	\int_{C(0)}^{C(t)} dt'e^{i\Delta_i^a(t' - t)}\sum_j n_if_{ji}(\alpha)\int_{C(0)}^{C(t')}dt''e^{i\Delta_j^a(t'' - t')}f_{aj}(\alpha)n_j(1 - n_a).
\end{equation}
The two terms in the expression for the grand potential are the two 3rd order diagrams (particle and hole-type diagrams) arising from the 3rd order skeleton depicted in Figure~\ref{fig:pt}(b).
Each term has 3 factors of $f$ which means that there will be 6 total terms that are 1st order in $\alpha$:
\begin{align}
	(1) &= -\frac{i}{\beta}\int_C dt \sum_{ia} i\alpha \delta_C(t - t_1)\delta_{ia}
	\int_{C(0)}^{C(t)} dt'e^{i\Delta_i^a(t' - t)}\sum_b (1 - n_a)f_{ab}
	\int_{C(0)}^{C(t')}dt''e^{i\Delta_i^b(t'' - t')}f_{bi}n_i(1 - n_b)\\
	(2) &= -\frac{i}{\beta}\int_C dt \sum_{ia} f_{ia}
	\int_{C(0)}^{C(t)} dt'e^{i\Delta_i^a(t' - t)}\sum_b (1 - n_a)i\alpha\delta_C(t' - t_1)\delta_{ab}
	\int_{C(0)}^{C(t')}dt''e^{i\Delta_i^b(t'' - t')}f_{bi}n_i(1 - n_b)\\
	(3) &= -\frac{i}{\beta}\int_C dt \sum_{ia} f_{ia}
	\int_{C(0)}^{C(t)} dt'e^{i\Delta_i^a(t' - t)}\sum_b (1 - n_a)f_{ab}
	\int_{C(0)}^{C(t')}dt''e^{i\Delta_i^b(t'' - t')}i\alpha\delta_C(t'' - t_1)\delta_{bi}n_i(1 - n_b)\\
	(4) &= \frac{i}{\beta}\int_C dt \sum_{ia} i\alpha \delta_C(t - t_1)\delta_{ia}
	\int_{C(0)}^{C(t)} dt'e^{i\Delta_i^a(t' - t)}\sum_j n_if_{ji}\int_{C(0)}^{C(t')}dt''e^{i\Delta_j^a(t'' - t')}f_{aj}n_j(1 - n_a)\\
	(5) &= \frac{i}{\beta}\int_C dt \sum_{ia} f_{ia}
	\int_{C(0)}^{C(t)} dt'e^{i\Delta_i^a(t' - t)}\sum_j n_ii\alpha\delta_C(t' - t_1)\delta_{ji}\int_{C(0)}^{C(t')}dt''e^{i\Delta_j^a(t'' - t')}f_{aj}n_j(1 - n_a)\\
	(6) &= \frac{i}{\beta}\int_C dt \sum_{ia} f_{ia}
	\int_{C(0)}^{C(t)} dt'e^{i\Delta_i^a(t' - t)}\sum_j n_if_{ji}\int_{C(0)}^{C(t')}dt''e^{i\Delta_j^a(t'' - t')}i\alpha\delta_C(t'' - t_1)\delta_{aj}n_j(1 - n_a)
\end{align}
We will now perform algebra similar to that of previous section to simplify terms (1)-(3):
\begin{align}
	(1) &= \frac{\alpha}{i\beta\Delta_p^b}\sum_{pb}f_{pb} f_{bp}
	(1 - n_p)n_p(1 - n_b)\left[t_1 + \frac{1}{i\Delta_p^b}\left(e^{-i\Delta_p^bt_1} - 1\right)\right]\\
	(2) &= \frac{\alpha}{\beta}\sum_{ia} f_{ia}f_{ai}(1 - n_a)n_i(1 - n_a)
	\frac{1}{(i\Delta_i^a)^2}\left[-e^{i\Delta_i^at_1}e^{-\beta\Delta_i^a} + e^{-\beta\Delta_i^a} + 1 -e^{-i\Delta_i^at_1}\right]\\
	(3) &= \frac{\alpha}{\beta}\sum_{pa} f_{pa}(1 - n_a)f_{ap}n_p(1 - n_p)
	\frac{1}{i\Delta_p^a}\left[-t_1 - \frac{1}{i\Delta_p^a} 
	-i\beta + \frac{e^{i\Delta_i^at_1}}{i\Delta_i^a}e^{-\beta\Delta_p^a}\right]
\end{align}
Terms (4)-(6) are analogous.

Examining these expressions, we see that each contains terms dependent on $t_1$ and that the prefactors of (1) and (3) are the same, but the prefactor of (2) is different. We will now use the notation $(1)[t_1]$ to indicate those terms in (1) that are dependent on $t_1$, and we will show that
\begin{equation}
	(1)[t_1] + (2)[t_1] + (3)[t_1] = 0.
\end{equation}
There is a slight subtlety here in that while we know that (1), (2), and (3) all have well-defined, finite, limits as $\Delta \rightarrow 0$, the same is not true of $(1)[t_1]$ for example. So, for the remainder of this section we will assume that $\Delta$ is finite and show that all the $t_1$ dependence cancels in this case. Since we know that $(1) + (2) + (3)$ has a finite limit, this should also be sufficient to show that the $t_1$-dependence vanishes in the limit as $\Delta \rightarrow 0$, but the expressions that we will be be working with will themselves not be valid in this limit. If we wanted to be perfectly rigorous, we would consider this limit separately.

The $t_1$-dependent parts are
\begin{align}
	(1)[t_1] &= \frac{\alpha}{\beta}\sum_{pa}f_{pa} f_{ap}
	(1 - n_p)n_p(1 - n_a)\frac{1}{i\Delta_p^a}\left[t_1 + 
	\frac{1}{i\Delta_p^a}e^{-i\Delta_p^at_1}\right] \\
	(2)[t_1] &= \frac{\alpha}{\beta}\sum_{ia} f_{ia}f_{ai}(1 - n_a)n_i(1 - n_a)
	\frac{1}{(i\Delta_i^a)^2}\left[-e^{i\Delta_i^at_1}e^{-\beta\Delta_i^a} -
	e^{-i\Delta_i^at_1}\right]\\
	(3)[t_1] &= \frac{\alpha}{\beta}\sum_{pa} f_{pa}(1 - n_a)f_{ap}n_p(1 - n_p)
	\frac{1}{i\Delta_p^a}\left[-t_1 + 
	\frac{e^{i\Delta_p^at_1}}{i\Delta_p^a}e^{-\beta\Delta_p^a}\right].
\end{align}
Each expression has two terms and the 1st term of $(1)[t_1]$ cancels the first term of $(3)[t_1]$ so that the sum of all of them is given by
\begin{align}
	(1)[t_1] + (2)[t_1] + (3)[t_1]&= \frac{\alpha}{\beta}\sum_{pa}f_{pa} f_{ap}
	(1 - n_p)n_p(1 - n_a)\frac{1}{(i\Delta_p^a)^2}e^{-i\Delta_p^at_1} \\
	&- \frac{\alpha}{\beta}\sum_{ia} f_{ia}f_{ai}(1 - n_a)n_i(1 - n_a)
	\frac{1}{(i\Delta_i^a)^2}e^{i\Delta_i^at_1}e^{-\beta\Delta_i^a}\\
	&- \frac{\alpha}{\beta}\sum_{ia} f_{ia}f_{ai}(1 - n_a)n_i(1 - n_a)
	\frac{1}{(i\Delta_i^a)^2}e^{-i\Delta_i^at_1}\\
	&+ \frac{\alpha}{\beta}\sum_{pa} f_{pa}(1 - n_a)f_{ap}n_p(1 - n_p)
	\frac{1}{(i\Delta_p^a)^2}e^{i\Delta_p^at_1}e^{-\beta\Delta_p^a}
\end{align}
A critical simplification is acheived by multiplying the real exponential factors into the occupation numbers so that, for example, 
\begin{equation}
	(1 - n_a)n_i(1 - n_a)e^{-\beta\Delta_i^a} = (1 - n_a)(1 - n_i)n_a.
\end{equation}
The results are that
\begin{align}
	(1)[t_1] + (2)[t_1] + (3)[t_1]&= \frac{\alpha}{\beta}\sum_{pa}f_{pa} f_{ap}
	(1 - n_p)n_p(1 - n_a)\frac{1}{(i\Delta_p^a)^2}e^{-i\Delta_p^at_1} \\
	&- \frac{\alpha}{\beta}\sum_{ia} f_{ia}f_{ai}(1 - n_a)(1 - n_i)n_a
	\frac{1}{(i\Delta_i^a)^2}e^{i\Delta_i^at_1}\\
	&- \frac{\alpha}{\beta}\sum_{ia} f_{ia}f_{ai}(1 - n_a)n_i(1 - n_a)
	\frac{1}{(i\Delta_i^a)^2}e^{-i\Delta_i^at_1}\\
	&+ \frac{\alpha}{\beta}\sum_{pa} f_{pa}f_{ap}n_a(1 - n_p)(1 - n_p)
	\frac{1}{(i\Delta_p^a)^2}e^{i\Delta_p^at_1}
\end{align}
We can now relabel indices so that in the second line $a,i \rightarrow p,a$ and in the 4th line $a,p \rightarrow i,a$:
\begin{align}
	(1)[t_1] + (2)[t_1] + (3)[t_1]&= \frac{\alpha}{\beta}\sum_{pa}f_{pa} f_{ap}
	(1 - n_p)n_p(1 - n_a)\frac{1}{(i\Delta_p^a)^2}e^{-i\Delta_p^at_1} \\
	&- \frac{\alpha}{\beta}\sum_{ap} f_{ap}f_{pa}(1 - n_p)(1 - n_a)n_p
	\frac{1}{(i\Delta_a^p)^2}e^{i\Delta_a^pt_1}\\
	&- \frac{\alpha}{\beta}\sum_{ia} f_{ia}f_{ai}(1 - n_a)n_i(1 - n_a)
	\frac{1}{(i\Delta_i^a)^2}e^{-i\Delta_i^at_1}\\
	&+ \frac{\alpha}{\beta}\sum_{ai} f_{ai}f_{ia}n_i(1 - n_a)(1 - n_a)
	\frac{1}{(i\Delta_a^i)^2}e^{i\Delta_a^it_1}
\end{align}
Everything is now close to canceling completely. We note that
\begin{equation}
	\Delta_p^q = -\Delta_q^p
\end{equation}
which allows us to write
\begin{align}
	(1)[t_1] + (2)[t_1] + (3)[t_1]&= \frac{\alpha}{\beta}\sum_{pa}f_{pa} f_{ap}
	(1 - n_p)n_p(1 - n_a)\frac{1}{(i\Delta_p^a)^2}e^{-i\Delta_p^at_1} \\
	&- \frac{\alpha}{\beta}\sum_{ap} f_{ap}f_{pa}(1 - n_p)n_p(1 - n_a)
	\frac{1}{(i\Delta_p^a)^2}e^{-i\Delta_p^at_1}\\
	&- \frac{\alpha}{\beta}\sum_{ia} f_{ia}f_{ai}(1 - n_a)n_i(1 - n_a)
	\frac{1}{(i\Delta_i^a)^2}e^{-i\Delta_i^at_1}\\
	&+ \frac{\alpha}{\beta}\sum_{ai} f_{ai}f_{ia}(1 - n_a)n_i(1 - n_a)
	\frac{1}{(i\Delta_i^a)^2}e^{-i\Delta_i^at_1}.
\end{align}
Finally, it is clear that the first line cancels the second and the 3rd line cancels the 4th. Thus, we can say that (1) + (2) + (3) makes a contribution to the particle number that is independent of $t_1$. An exactly analogous argument can presumably be made for (4) + (5) + (6). This means that for 3rd order perturbation theory each of the two 3rd order diagrams individually conserves particle number.
\end{widetext}

\subsection{Some features of 4th order perturbation theory}
4th order perturbation theory becomes quite tedious, but it is sufficient for our purposes to consider the pair of 4th order diagrams that arise from the quadratic term in the CCS equations. These diagrams correspond to the skeleton listed 3rd in Figure~\ref{fig:pt}(c):
\begin{widetext}
\begin{align}
	\Omega^{[4]\ast}(\alpha) &= \frac{i}{\beta}\int_Cdt \sum_{ia}
	f_{ia}(\alpha) s_i^a(t)^{[3\ast]} \nonumber \\
	&= \frac{i}{\beta}\int_Cdt \sum_{ia}
	f_{ia}(\alpha) (-i)\int_{C(0)}^{C(t)}dt' e^{i\Delta_i^a(t' - t)}
	\sum_{jb} f_{jb}(\alpha)s_j^a(t')^{[1]}s_i^b(t')^{[1]}
\end{align}
We have indicated this particular pair of 4th order contributions by an asterisk. Plugging in the definition of the 1st order $s$ amplitudes given in Equation~\ref{eqn:s1} yields:
\begin{align}
	\Omega^{[4]\ast}(\alpha) &=\frac{1}{\beta}\int_Cdt \sum_{ia}
	f_{ia}(\alpha) \int_{C(0)}^{C(t)}dt' e^{i\Delta_i^a(t' - t)}
	\sum_{jb}f_{jb}(\alpha)\nonumber \\
	&\times (-i)\int_{C(0)}^{C(t')}dt''e^{i\Delta_j^a(t'' - t')}f_{aj}(\alpha)n_j(1 - n_a) (-i)\int_{C(0)}^{C(t')}dt''e^{i\Delta_i^b(t'' - t')}f_{bi}(\alpha)n_i(1 - n_b)
\end{align}
Just as before, we expand this expression as a function of $\alpha$ to obtain
\begin{equation}
	\Omega^{[4]\ast}(\alpha) = \Omega^{[4]\ast}_0 + \Omega^{[4]\ast}_1\alpha + \frac{1}{2}\Omega^{[4]\ast}_2 \alpha^2 + \ldots 
\end{equation}

We are interested in the coefficient of $\alpha$ in this expansion since it gives the 4th order ($\ast$) correction to the particle number. Clearly there are 4 terms:
\begin{align}
	(1) &=\frac{1}{\beta}\int_Cdt \sum_{ia}
	\delta_{ia}\delta_C(t - t_1) \int_{C(0)}^{C(t)}dt' e^{i\Delta_i^a(t' - t)}
	\sum_{jb}f_{jb}\nonumber \\
	&\times (-i)\int_{C(0)}^{C(t')}dt''e^{i\Delta_j^a(t'' - t')}f_{aj}n_j(1 - n_a) (-i)\int_{C(0)}^{C(t')}dt''e^{i\Delta_i^b(t'' - t')}f_{bi}n_i(1 - n_b) \\
	(2) &=\frac{1}{\beta}\int_Cdt \sum_{ia}
	f_{ia} \int_{C(0)}^{C(t)}dt' e^{i\Delta_i^a(t' - t)}
	\sum_{jb}\delta_{jb}\delta_C(t' - t_1)\nonumber \\
	&\times (-i)\int_{C(0)}^{C(t')}dt''e^{i\Delta_j^a(t'' - t')}f_{aj}n_j(1 - n_a) (-i)\int_{C(0)}^{C(t')}dt''e^{i\Delta_i^b(t'' - t')}f_{bi}n_i(1 - n_b)
	\\
	(3) &=\frac{1}{\beta}\int_Cdt \sum_{ia}
	f_{ia} \int_{C(0)}^{C(t)}dt' e^{i\Delta_i^a(t' - t)}
	\sum_{jb}f_{jb}\nonumber \\
	&\times (-i)\int_{C(0)}^{C(t')}dt''e^{i\Delta_j^a(t'' - t')}\delta_{aj}\delta_C(t'' - t_1)n_j(1 - n_a) (-i)\int_{C(0)}^{C(t')}dt''e^{i\Delta_i^b(t'' - t')}f_{bi}n_i(1 - n_b) \\
	(4) &=\frac{1}{\beta}\int_Cdt \sum_{ia}
	f_{ia}\int_{C(0)}^{C(t)}dt' e^{i\Delta_i^a(t' - t)}
	\sum_{jb}f_{jb}\nonumber \\
	&\times (-i)\int_{C(0)}^{C(t')}dt''e^{i\Delta_j^a(t'' - t')}f_{aj}n_j(1 - n_a) (-i)\int_{C(0)}^{C(t')}dt''e^{i\Delta_i^b(t'' - t')}\delta_{bi}\delta_C(t'' - t_1)n_i(1 - n_b)
\end{align}

After a great deal of algebra, we obtain
\begin{align}
	(1) &=-\frac{1}{\beta}\sum_{pjb}n_pn_j(1 - n_p)(1 - n_b)f_{jb}f_{bp}f_{pj}
	\frac{1}{i\Delta_j^bi\Delta_p^b}\nonumber \\
	&\times\left[t_1 + \frac{1}{i\Delta_j^p}\left(e^{-i\Delta_j^pt_1} - 1\right)  + \frac{1}{i\Delta_p^b}\left(e^{-i\Delta_p^bt_1} - 1\right) - \frac{1}{i\Delta_j^b}\left( e^{-i\Delta_j^bt_1} - 1\right)\right]\\
	(2) &=-\frac{1}{\beta}\sum_{iap}(1 - n_i)n_pn_a(1 - n_p)f_{ia}f_{ap}f_{pi}
	\frac{1}{i\Delta_p^a}
	\frac{1}{i\Delta_i^p}\frac{1}{i\Delta_i^a}
	\left[1 - e^{-i\Delta_p^it_1} - e^{-i\Delta_a^pt_1} + e^{i\Delta_i^at_1}\right] \nonumber \\
	&-\frac{1}{\beta}\sum_{iap}n_in_p(1 - n_a)(1 - n_p)f_{ia}f_{ap}f_{pi}
	\frac{1}{i\Delta_p^a}\frac{1}{i\Delta_i^p}\frac{1}{i\Delta_i^a}
	\left[e^{-\Delta_i^at_1} - e^{-i\Delta_p^at_1} - e^{-i\Delta_i^pt_1} + 1\right]\\
	(3) &=-\frac{1}{\beta}\sum_{ipb}n_in_p(1 - n_p)(1 - n_b)f_{bi}f_{ip}f_{pb}
	\frac{1}{i\Delta_i^b}\nonumber \\ 
	&\times \left\{\frac{1}{i\Delta_i^p}\left[-i\beta - t_1 - \frac{1}{i\Delta_i^p}\left(e^{i\Delta_i^pt_1}e^{-\Delta_i^p\beta)} - 1\right)\right] - \frac{1}{i\Delta_b^p}\left[\frac{1}{i\Delta_b^i}\left(e^{\beta\Delta_b^i} - e^{i\Delta_b^it_1}\right) - \frac{1}{i\Delta_i^p}\left(e^{i\Delta_b^pt_1}e^{-\beta\Delta_i^p} - 1\right)\right]\right\}\\
	(4) &=-\frac{1}{\beta}\sum_{jap}n_pn_j(1 - n_a)(1 - n_p)f_{aj}f_{jp}f_{pa}
	\frac{1}{i\Delta_j^a}\nonumber\\
	&\times \left\{\frac{1}{i\Delta_p^a}\left[-i\beta - t_1 + \frac{1}{i\Delta_p^a}\left(e^{-\beta\Delta_p^a}e^{i\Delta_p^at_1} - 1\right)\right]  
	- \frac{1}{i\Delta_j^p}\left[\frac{1}{i\Delta_j^a}\left(e^{-\beta\Delta_j^a} - e^{-i\Delta_j^at_1}\right) - \frac{1}{i\Delta_j^p}\left(e^{-\beta\Delta_p^a}e^{-i\Delta_j^pt_1} - 1\right)\right]\right\}
\end{align}
These expressions are complicated and difficult to verify, but it is sufficient to examine the coefficients of the terms linear in $t_1$ to see that for (1) + (2) + (3) + (4), the $t_1$-dependence does not cancel. Since the sum of all 4th order terms should have no dependence on $t_1$ as per the argument in Section~\ref{sec:KelConserv}, this also implies that $\Omega^{[4]} - \Omega^{[4]\ast}$ does not generate dynamics that conserve particle number. Since LCCS includes the $\Omega^{[4]} - \Omega^{[4]\ast}$ contribution at 4th order, this suggests that Keldysh-LCCS will not conserve particle number.
\end{widetext}

\subsection{Numerical results}\label{sec:Anumeric}
From this discussion and the discussion given in Section~\ref{sec:KelConserv}, we conclude several things about particle number conservation in the 1-particle problem:
\begin{enumerate}
	\item $n$th order perturbation theory should conserve particle number
	\item For 2nd and 3rd order perturbation theory, each diagram {\it individually} conserves the particle number
	\item But this is not true of 4th order perturbation theory where the ($\ast$) terms (and therefore the remaining terms also) do not conserve particle number
	\item LCCS does not conserve particle number
	\item CCS does conserve particle number by virtue of being exact: it is complete to all orders in perturbation theory.
\end{enumerate}
We will now verify numerically these statements. Because there is error in the particle number due to the numerical integration, we will say that a given theory does not conserve particle number if the quantity changes in time and this change does not decrease as the grid is made finer. Our example Hamiltonian is a two-level system,
\begin{equation}
	H = \begin{pmatrix}
	E_1 & 0 \\
	0 & E_2
	\end{pmatrix} + 
	\begin{pmatrix}
	V_{11} & V_{12} \\
	V_{21} & V_{22} \\
	\end{pmatrix}.
\end{equation}
For this example, we use
\begin{equation}
	E_1 = 0.1 \qquad E_2 = 0.4
\end{equation}
\begin{equation}
	V_{11} = V_{22} = 0.1 \qquad V_{12} = V_{21}^{\ast} = 1 + i/2.
\end{equation}
The relatively large perturbation is used to accentuate the defects in the theory. We use $\mu = 0.0$ and $T = 0.5$ in these calculations, and compute $N$ at points between $t = 0$ and $t = 1$. For the exact dynamics of the system, $\avg{N}$ is conserved. 

In Table~\ref{tab:N}, we show the number of particles for different approximate methods for the equilibrium grand potential. As expected, different many-body methods will lead to different numbers of electrons at finite temperature, and CCS is exact for this 1-particle problem.
\begin{table}
\begin{tabular}{c|ccc}
\hline\hline
method & $N$ & error& \% relative error\\
\hline
PT2 & 0.6679063& -0.20734& 23.69\\
PT3 & 0.9500828& 0.07484& 8.55\\
PT4 & 1.0446668& 0.16942& 19.36\\
PT4$\ast$ & 1.0580504& 0.18281& 20.89\\
LCCS & 0.9436346& 0.06839& 7.81\\
CCS & 0.8752419& -3.6189E-07& 4.13E-05\\
exact& 0.8752423& -& -\\
\hline\hline
\end{tabular}
\caption{\label{tab:N} The average number of particles, $\avg{N}$, for each method in equilibrium. Note that the small error in CCS is due to the numerical integration and to error in the finite difference determination of $\avg{N}$.}
\end{table}

\begin{figure}
\includegraphics[scale=1.0]{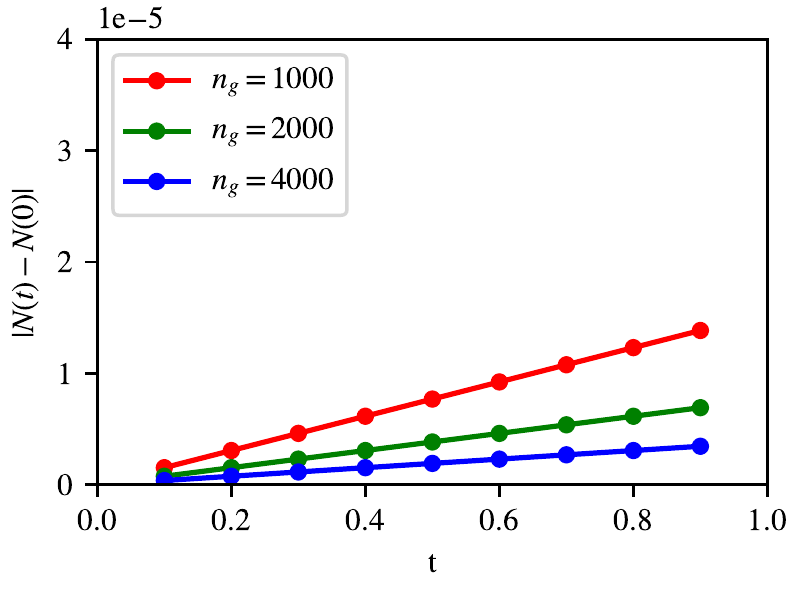}
\caption{\label{fig:mp2} Absolute difference in particle number relative to the particle number at $t = 0$ as a function of time for Keldysh PT at 2nd order. Note the $y$-axis is scaled by 1e-5.}
\end{figure}

\begin{figure}
\includegraphics[scale=1.0]{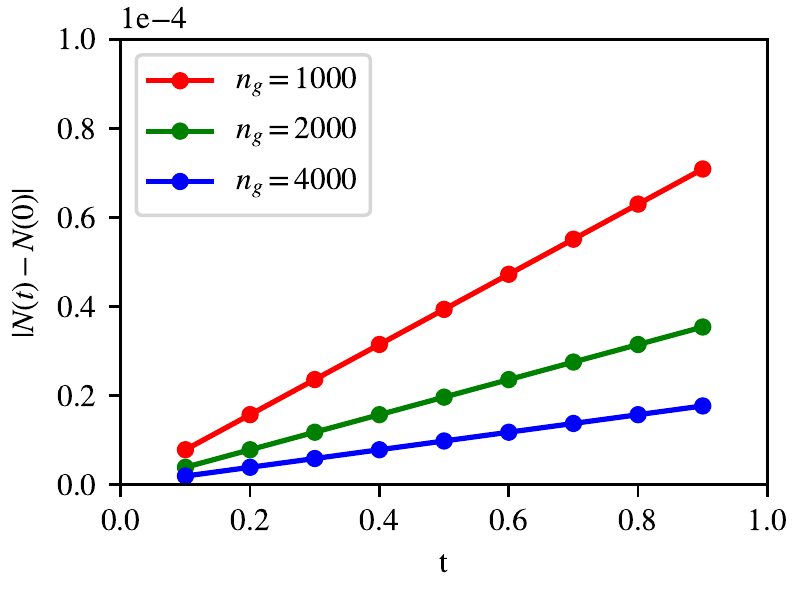}
\caption{\label{fig:mp3} Absolute difference in particle number relative to the particle number at $t = 0$ as a function of time for Keldysh PT at 3rd order. Note the $y$-axis is scaled by 1e-4.}
\end{figure}

\begin{figure}
\includegraphics[scale=1.0]{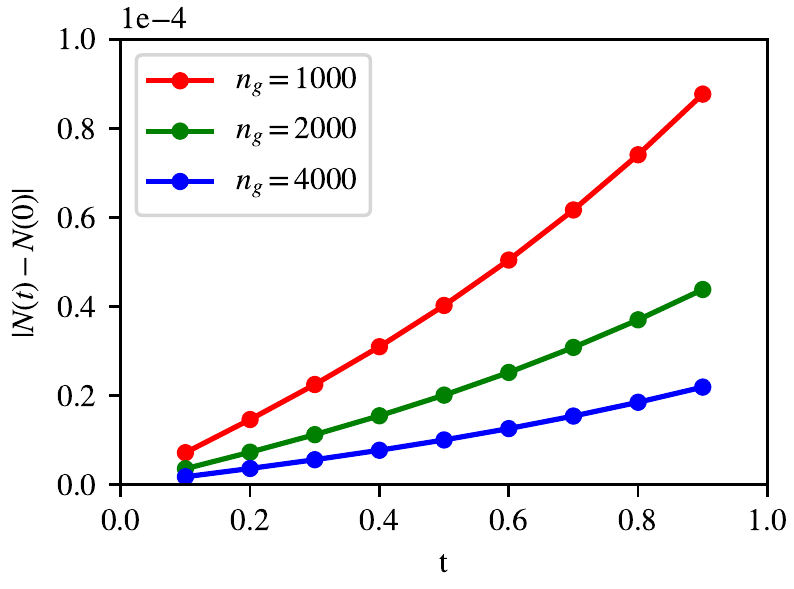}
\caption{\label{fig:mp4} Absolute difference in particle number relative to the particle number at $t = 0$ as a function of time for Keldysh PT at 4th order. Note the $y$-axis is scaled by 1e-4.}
\end{figure}

In the following time-dependent calculations, separate calculations are performed at each time point and the property of interest, $N$, is computed by finite difference differentiation. Since separate calculations have to be performed for each time point anyway, we use the same number of grid points at each time point. This leads to an integration error which is easy to recognize and separate from errors in the many-body approximations themselves. In Figures~\ref{fig:mp2}-\ref{fig:mp4} we confirm numerically point (1).
These figures indicate that for perturbation theory at any order, $\avg{N}$ will be conserved as a function of time in the limit of exact numerical integration.

\begin{figure}
\includegraphics[scale=1.0]{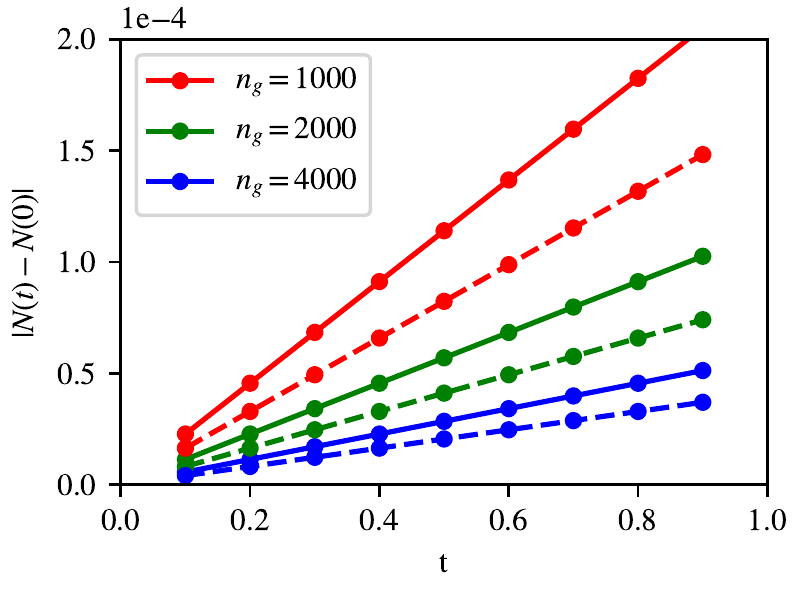}
\caption{\label{fig:mp3ph} Absolute difference in particle number relative to the particle number at $t = 0$ as a function of time for each contribution to Keldysh PT at 3rd order. The solid lines show the contribution from the ``particle'' diagram and the dotted lines show the contribution from the ``hole'' diagram. Both appear to individually conserve particle number.}
\end{figure}

Point (2) is trivially true at 2nd order where there is only a single diagrammatic contribution. At third order however, there are two diagrams which we refer to as  the ``particle'' and ``hole'' diagrams for the diagrams containing more particle or hole propagators respectively. In Figure~\ref{fig:mp3ph} we show that each of these contributions individually conserves $N$.
This confirms numerically the result of Section~\ref{sec:mp3}.

\begin{figure}
\includegraphics[scale=1.0]{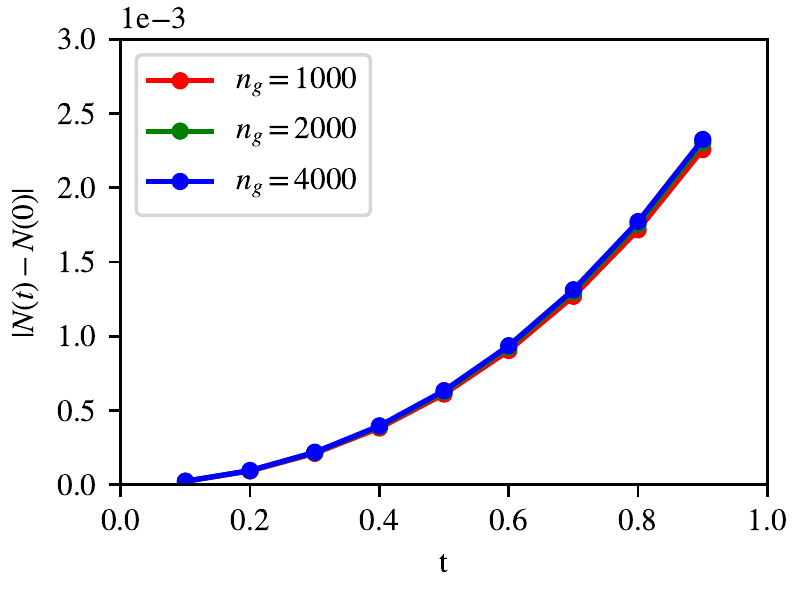}
\caption{\label{fig:mp4s} Absolute difference in particle number relative to $\avg{N}$ at $t = 0$ as a function of time for Keldysh PT at 3rd order including the 4th order $\ast$ terms. Note that increasing the number of grid points does not decrease the difference in $N(t)$; it even increases slightly. Also the differences are about an order of magnitude larger than those seen in the previous examples.}
\end{figure}

Point (3) is confirmed numerically in Figure~\ref{fig:mp4s} where the $\ast$ contributions at 4th order are included and the particle number is no longer conserved.

\begin{figure}
\includegraphics[scale=1.0]{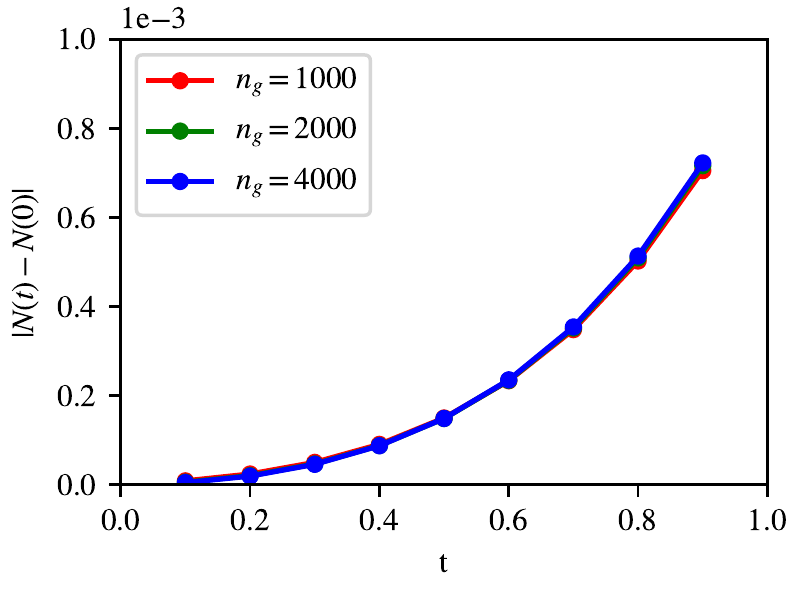}
\caption{\label{fig:lccs} Absolute difference in particle number relative to $\avg{N}$ at $t = 0$ as a function of time for LCCS. Note that increasing the number of grid points does not decrease the difference in $N(t)$.}
\end{figure}

Point(4) is confirmed in Figure~\ref{fig:lccs} which qualitatively resembles Figure~\ref{fig:mp4s}. 

\begin{figure}
\includegraphics[scale=1.0]{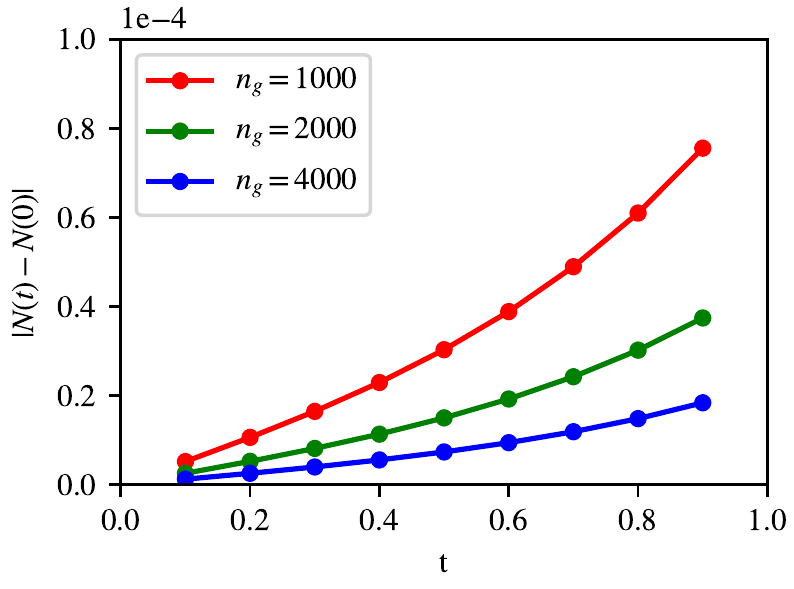}
\caption{\label{fig:ccs} Absolute difference in particle number relative to the value at $t = 0$ as a function of time for CCS. $N$ appears to be conserved as we would expect from a theory that is exact.}
\end{figure}

Finally, point (5) is corroborated in Figure~\ref{fig:ccs}.

In this particular example, the fact that particle number is not conserved does not create a significant problem because the differences on this time-scale are small compared to the error in the various methods at equilibrium (see Table~\ref{tab:N}), but this will not be true in general.

In this Appendix, we have examined particle-number conservation of different perturbation theory and coupled cluster models for finite-temperature dynamics without two-particle interactions. The problem is trivially solvable, but it nonetheless allows us to show explicitly conclusions (1) - (5). In Section~\ref{sec:Anumeric} we verified these results numerically. This discussion suggests that in more general Keldysh-CC models, such as the Keldysh-CCSD presented in this article, global symmetries will not be preserved by the dynamics. The seriousness of this flaw will, in general, depend on the system.

\bibliography{References}
\end{document}